\documentclass[10pts]{article}
\usepackage{amsthm}
\theoremstyle{plain}
\usepackage{graphicx} 
\usepackage{braket}
\usepackage{amsmath}
\usepackage{epstopdf}

\newtheorem*{theorem*}{Theorem}
\usepackage{amsfonts}
\begin{document}

\title{\bf A One-Particle Time of Arrival Operator for a Free Relativistic Spin-$0$ Charged Particle in $(1+1)$ Dimensions}
\author{Joseph Bunao and Eric A. Galapon\thanks{eagalapon@up.edu.ph, eric.galapon@upd.edu.ph}  \\ Theoretical Physics Group, National Institute of Physics\\University of the Philippines, 1101 Philippines}
\maketitle
\begin{abstract}
We construct a one-particle TOA operator $\mathcal{\hat{T}}$ canonically conjugate with the Hamiltonian describing a free, charged, spin-$0$, relativistic particle in one spatial dimension and show that it is maximally symmetric. We solve for its eigenfunctions and show that they form a complete and non-orthogonal set. Plotting the time evolution of their corresponding probability densities, it implies that the eigenfunctions become more localized at the origin at the time equal to their eigenvalues. That is, a particle being described by an eigenfunction of $\mathcal{\hat{T}}$ is in a state of definite arrival time at the origin and at the corresponding eigenvalue. We also calculate the TOA probability distribution of a particle in some initial state.
\end{abstract}

\section{Introduction}

The notion of time is one of the problems obstructing the marriage of Einstein's General Theory of Relativity and Standard Quantum Mechanics into one framework of Quantum Gravity \cite{qgart}. The two theories have a mutually incompatible treatment of time. For general relativity, time has a dynamic and intrinsic role in the evolution of the system being studied. For quantum mechanics, however, time is merely an extrinsic parameter marking the evolution of the system. The system does not affect it nor does it affect the system. This pessimistic view on time was prominent in the earlier days of quantum mechanics. The prevalent formulation was the von Neumann (standard) formulation of quantum mechanics which implied that time is not a dynamic physical entity to be measured - time, according to that view, is not an observable. Observables in quantum mechanics are represented by mathematical objects called operators and in the standard formulation, these operators should be self-adjoint. The demotion of time into just being a parameter stems from Pauli's theorem which says that there can be no self-adjoint time operator canonically conjugate with its corresponding semi-bounded system Hamiltonian \cite{pauli}. This pessimism towards time was apparent through the trilogy of papers by Allcock \cite{allc1,allc2,allc3} which suggests that one cannot find a distribution for the time it takes for a free particle to arrive at a certain point - the particle's Time of Arrival (TOA).

This pessimism seems misguided however, since we can, for example, measure the TOA of a particle. As an answer to this, studies to promote the role of time has been made in recent years. In the framework of Positive Operator Valued Measures (POVMs), the standard formulation of quantum mechanics can be extended by saying that quantum observables are not necessarily self-adjoint operators but may be non-self-adjoint, maximally-symmetric operators which are the first operator moments of POVMs. This framework lets one bypass Pauli's theorem by allowing one to consider non-self-adjoint, maximally-symmetric time operators as quantum observables. Moreover in \cite{galpauli}, it was rigorously shown that the assumption of the existence of a bounded, self-adjoint time operator canonically conjugate to a semi-bounded Hamiltonian is consistent. It was seen that, in the steps followed by Pauli, there were some implicit assumptions made and that these assumptions were inconsistent. It is then the case that Pauli's theorem simply does not hold in the standard formulation of quantum mechanics. This then opens up an avenue to still consider time as an observable in standard quantum mechanics. In line with this optimistic view, several studies on TOA operators 
\cite{abohm, grot, gal1, gal2, gal3, gal4, gal5, gal6, sombillo, galmuga, muga1, muga2, muga3} 
were made. Specifically in \cite{gal4}, it was seen that a time operator, conjugate with its corresponding Hamiltonian, can be written as a sum of the Bender-Dunne basis operators $\hat{T}_{m,n}$ \cite{ben1, ben2}. As an example, the time of arrival for a free non-relativistic particle in one spatial dimension would then be given by $-m\hat{T}_{-1,1}$ \cite{abohm, gal4}, where $m$ is the mass of the particle. Some applications of these TOA operators would be a test to the possibly inconsistent assumptions of neutron time-of-flight spectroscopy experiments \cite{gal7} (that is, simultaneously assuming the validity of the classical time of arrival and the quantum mechanical broad de Broglie wave packets), the delay in photoemission from atoms which is in conflict with the assumption of spontaneous photoemission \cite{delayp}, and the upper limit to a quantum tunneling delay time being much shorter than previously predicted \cite{tundel}.

In our quest to promote time as an observable, that is, to make time be on equal footing as the other physical quantities, the next system to study time in would be the quantum mechanics of a relativistic particle 
\cite{oth8, relnonrel, intop, velop, dir}. 
In this regime, special relativity becomes relevant and it tells us that space and time should be treated on the same ground. If position is to be taken as an observable of the particle, it is then natural to promote time (such as the TOA at a certain location) also as an observable with a corresponding operator. This TOA operator should then be conjugate with the system Hamiltonian which may be derived from the equations of motion describing the system. Another property of the relativistic quantum particle is its spin. Particles with different spins are described by different equations of motion. Some of the well known equations of motion are the Klein-Gordon equation, describing spin-$0$ particles, and the Dirac equation, describing spin-$1/2$ particles. However, there are certain problems arising in these equations, problems such as 'negative energies' and 'negative probabilities'. These can be attributed to the existence of anti-particles. They are just particles with the same mass but opposite charge of their corresponding particle pair. The particle anti-particle pairs are spontaneously created and annihilated in a relativistic quantum system so that the system does not have a fixed number of particles. This should not be too surprising since quantum mechanics tells us that energy can have quantum fluctuations and special relativity tells us that mass is just another form of energy, then it is natural that relativistic quantum systems may have a fluctuating number of massive particles. Standard quantum mechanics, however, was not formulated with a changing number of particles in a system. Quantum Field Theory (QFT), the theory of quantum fields where space and time are just labels of a point on the field, remedies this by saying that particles and anti-particles are just excitations of the quantum field which may undergo random quantum fluctuations, thus the creation and annihilation of particles and anti-particles. This is one of the strengths of QFT. We can, however, still ask the question: Given a relativistic particle with a certain initial state, what is the TOA probability distribution of that same particle? If we are to perform this TOA experiment, we would not be sure if the particle that arrived at our detector is the same one we started with. But surely, we need not the full machinery of QFT to answer the quantum mechanics of this relativistic particle. We just need to restrict our analysis to one particle \cite{book}. That is, we would only consider the cases where it is the same particle that has arrived at the detector of our TOA experiment.

The purpose of this paper then is to construct a TOA operator for a free, relativistic, spin-$0$, charged particle in one spatial dimension by solving for an operator $\mathcal{\hat{T}}$ conjugate with the system Hamiltonian derived from the Klein-Gordon equation. We are also explicit in making the constructed TOA operator a true one-particle operator. Other papers have also studied time operators in a relativistic context. These studies include \cite{oth8} which has also constructed a relativistic TOA operator for spin-$0$ particles which closely mirrors \cite{grot}. That is, the time evolution of the position operator was inverted to obtain an expression for a TOA operator defined with a suitable operator ordering. In \cite{relnonrel}, however, a kinematic time operator was defined for relativistic quantum mechanics but the dynamics suggest that the eigenstates of this time operator is unphysical. To remedy this, the Bohmian interpretation was invoked.
While in \cite{intop}, a self-adjoint time operator was constructed for a free massless relativistic particle in terms of the Poincare group generators. Under Lorentz boosts, the transformation of the operator differs from its expected transformation law from special relativity. This suggests that the concept of time associated with this operator differs from the Minkowski time coordinate. Another is in \cite{velop}, where an alternative set of coordinates are considered to avoid the false result of having the eigenvalues of the velocity operator of a massive relativistic particle be $\pm c$ which suggest that particles with mass travel at the speed of light. The corresponding operator of the 'zeroth' component of these new coordinates does not commute with the Hamiltonian of the system which gives us a time-energy uncertainty relation. In \cite{dir}, a self-adjoint dynamic time operator for relativistic spin-$1/2$ particles was constructed. The resulting commutation relation with the system Hamiltonian is analogous with the position-momentum commutation relations. Such relativistic time operators would be useful in atomic fountain clocks where they lower the temperature of the particles to hinder relativistic effects \cite{clock}. These procedures would no longer be needed if one can study the quantum behavior of time in a relativistic regime. One can see then that there are various approaches in the study of time in relativistic quantum mechanics. The hope is that upon further study, these approaches would give us a more refined understanding of time. Moreover, promoting time to be on the same footing as other physical observables may give us the insight to better understand the problem of time in Quantum Gravity \cite{qgart}. 

In section \ref{bgrdkg}, we first review the one-particle interpretation of the Klein-Gordon equation by closely following the discussions from \cite{book}. Here, we review how we can separate wavefunctions to describe either a positively charged or negatively charged particle, allowing us to have a non-negative probability density of the charged particle and a notion of true one-particle operators. We construct the TOA operator by solving its canonical commutation relation with the Hamiltonian of the Klein-Gordon particle in section \ref{constr}, make it into a true one-particle operator (which we will later call as $\mathcal{\hat{T}}$) in section \ref{onep}, and show that it is symmetric in section \ref{symmop}. We solve for the eigenfunctions of $\mathcal{\hat{T}}$ in section \ref{eigfsec}, show that they form a complete and non-orthogonal set in sections \ref{complete} and \ref{nonorthsec}, respectively, and study their dynamical behaviour by plotting the time evolution of their associated probability densities in section \ref{dynamics}. In section \ref{pdist}, we calculate the probability (density) that the particle in some initial state will arrive at the origin at time $\tau$ and, as an example, considered the initial state to be normally distributed about some initial position and momentum. Lastly in section \ref{conc}, we conclude.

\section{One-Particle Interpretation of Klein-Gordon Particles}
\label{bgrdkg}
In one spatial dimension, the wavefunction $\psi$ of a free, relativistic, spin-$0$ particle with mass $m_{0}$ and magnitude of charge $e$ is described by the Klein-Gordon Equation (KGE)
\begin{equation}\label{kge}
\frac{1}{c^{2}}\frac{\partial^{2} \psi}{\partial t^{2}} - \frac{\partial^{2} \psi}{\partial x^{2}} + \frac{m_{0}^{2}c^{2}}{\hbar^{2}}\psi = 0.
\end{equation}
Along with Eq. (\ref{kge}), a $\rq$probability density$\rq$ $\tilde{\rho}$ can be constructed
\begin{equation}\label{rhot}
\tilde{\rho} = \frac{i\hbar}{2m_{0}c^{2}}\left(\psi^{*}\frac{\partial \psi}{\partial t} - \psi\frac{\partial \psi^{*}}{\partial t}\right)
\end{equation}
However, since the KGE is second order in time, both $\psi$ and $\partial \psi/\partial t$ can have any arbitrary value at some given time. This means that $\tilde{\rho}$ is not positive definite and hence, not a probability density. This can be remedied by interpreting it as a charge density. That is, $\rho = e\tilde{\rho}$. In this interpretation, $\rho$ can then be allowed to be positive or negative as it just measures the difference between the number of positively charged and negatively charged particles. One sees then that there is another degree of freedom in $\psi$ corresponding to states describing positively and negatively charged particles. To be more concrete, consider the plane wave solutions for a particle with definite momentum $p$ which has the form $\tilde{\psi} \propto \exp\left(ipx/\hbar - iEt/\hbar\right)$. Then substituting $\tilde{\psi}$ back into Eq. (\ref{kge}), we get $E = \pm \sqrt{p^{2}c^{2}+m_{0}^{2}c^{4}} = \pm E_{p} = \lambda E_{p}$
where, $\lambda = \pm 1$ and $E_{p} = \sqrt{p^{2}c^{2}+m_{0}^{2}c^{4}}$, so that for a particle with a definite momentum $p$, we get two $\rq$energies$\rq$ giving rise to two different plane wave states. We can then characterize these plane wave states by the sign of $\lambda$. Explicitly, the plane wave states are $\tilde{\psi}_{\lambda = \pm 1} \propto \exp\left(ipx/\hbar - \lambda iE_{p}t/\hbar\right)$.
Moreover, the corresponding charge density $\rho = e\tilde{\rho}$ of $\tilde{\psi}_{\lambda = +1}$ ($\tilde{\psi}_{\lambda = -1}$) is positive (negative) definite. We can then interpret that $\tilde{\psi}_{\lambda = +1}$ ($\tilde{\psi}_{\lambda = -1}$) describes a positively (negatively) charged particle. Note that $E = \pm E_{p} = \lambda E_{p}$ does not imply that there is a positive and negative energy. The sign of $\lambda$ simply characterizes whether the state describes either a positively charged or negatively charged particle. 


However, as presented in Eq. (\ref{kge}), the KGE does not make this separation of solutions into positive and negative states obvious. It is best to make the charge degree of freedom more visible by decomposing the second-order-in-time KGE (Eq. (\ref{kge})) into two first-order-in-time differential equations. Moreover, this allows us to define a Hamiltonian operator for the system. The first natural step is to let $\psi$ and $\partial \psi/\partial t$ to be independent functions. Explicitly from \cite{book}, we let $\psi = \varphi + \chi$ and $i\hbar \frac{\partial \psi}{\partial t} =  m_{0}c^{2}(\varphi-\chi)$
so that Eq. (\ref{kge}) becomes separated into two equations made compact by a two-element column vector $\Psi = (\varphi \; \chi)^{T}$.
\begin{align}\label{kgesch}
i\hbar \frac{\partial \Psi}{\partial t} &= (\sigma_{3}+i\sigma_{2})\left(-\frac{\hbar^{2}}{2m_{0}}\right)\frac{\partial^{2} }{\partial x^{2}}\Psi + \sigma_{3}m_{0}c^{2}\Psi \equiv \mathcal{\hat{H}}_{\Psi}\Psi 
\end{align}
where, $\mathcal{\hat{H}}_{\Psi}$ is the Hamiltonian operator of the system, and
\begin{equation}\label{taus}
\sigma_{0} = 
\begin{pmatrix}
 1 & 0\\
 0 & 1 \\
\end{pmatrix},
\sigma_{1} = 
\begin{pmatrix}
 0 & 1\\
 1 & 0 \\
\end{pmatrix},
\sigma_{2} = 
\begin{pmatrix}
 0 & -i\\
 i & 0 \\
\end{pmatrix},
\sigma_{3} = 
\begin{pmatrix}
 1 & 0\\
 0 & -1 \\
\end{pmatrix}
\end{equation}
Note that $\sigma_{0}$ is just the identity matrix. We will follow \cite{book} and call Eq. (\ref{kgesch}) the Schrodinger representation of Eq. (\ref{kge}), or the $\Psi$-representation. The charge density can also be written as 
\begin{equation}\label{denpsi}
\rho = e\tilde{\rho} = e(\varphi^{*}\varphi - \chi^{*}\chi) = e\Psi^{\dagger}\sigma_{3}\Psi
\end{equation}
Eq. (\ref{kgesch}), however, is still a coupled differential equation and we can decouple it by diagonalizing $\mathcal{\hat{H}}_{\Psi}$. In the case we are investigating, the non-interacting case, this is possible. Firstly, it is more convenient to work in the momentum representation ($p$-representation) so that we make the replacement $-i\hbar\partial/\partial x \rightarrow p$ in the Hamiltonian $\mathcal{\hat{H}}_{\Psi}$. The functions from this point will also be in the $p$-representation unless otherwise stated. Then from \cite{art,book}, we introduce
\begin{equation}\label{u}
U^{\pm 1} = \frac{(m_{0}c^{2}+E_{p})\sigma_{0} \mp (m_{0}c^{2}-E_{p})\sigma_{1}}{\sqrt{4m_{0}c^{2}E_{p}}}
\end{equation}
where $U^{-1}$ is the inverse of $U$. The KGE (Eq. (\ref{kgesch})) then takes the form
\begin{align}\label{kgephi}
i\hbar \frac{\partial }{\partial t}\Phi = U\mathcal{\hat{H}}_{\Psi}U^{-1}\Phi = E_{p}\sigma_{3} \Phi \equiv \mathcal{\hat{H}}_{\Phi}\Phi
\end{align}
where $\Phi = U \Psi \equiv (\phi_{+}\;\phi_{-})^{T}$ is a two-element column vector 
and $\mathcal{\hat{H}}_{\Phi}$ is the corresponding Hamiltonian. Following \cite{book}, we call this representation the Feshbach-Villars representation or $\Phi$-representation for short. Also, if we want to transform an arbitrary operator $\mathcal{\hat{A}}_{\Psi}$ from the $\Psi$-representation to the $\Phi$-representation, we simply use the transform $\mathcal{\hat{A}}_{\Phi} = U\mathcal{\hat{A}}_{\Psi}U^{-1} $. We sometimes indicate $\Phi-p$ representation to emphasize that we are also in the momentum representation. Note that since $\mathcal{\hat{H}}_{\Phi}$ is diagonal, Eq. (\ref{kgephi}) does not mix $\phi_{+}$ and $\phi_{-}$. That is, the free time evolution of $\phi_{+}$ ($\phi_{-}$) depends only on $\phi_{+}$ ($\phi_{-}$) itself. To be more concrete, if we have $\Phi_{+} = (\phi_{+}\;0)^{T}$ and $\Phi_{-} = (0\;\phi_{-})^{T}$ then Eq. (\ref{kgephi}) becomes
\begin{align}
i\hbar \frac{\partial \phi_{+}}{\partial t} &= E_{p}\phi_{+} \label{kgephipl}\\
i\hbar \frac{\partial \phi_{-}}{\partial t} &= -E_{p}\phi_{-} \label{kgephim}\\
\nonumber
\end{align}
Moreover, we can calculate the charge density (albeit, density in momentum space in our present discussion). Following from Eq. (\ref{denpsi})
\begin{align}\label{densphi}
\rho = e\Psi^{\dagger}\sigma_{3}\Psi = e\Phi^{\dagger}\sigma_{3}\Phi = e\left(|\phi_{+}|^{2} - |\phi_{-}|^{2}\right)
\end{align}
so that the corresponding charge density for $\Phi_{+}$ ($\Phi_{-}$) is $e|\phi_{+}|^{2}$ ($-e|\phi_{-}|^{2}$). That is, $\Phi_{+}$ ($\Phi_{-}$) represents a positively (negatively) charged particle. One sees then more explicitly that had we started with a free positively (negatively) charged particle, then after its evolution given by Eq. (\ref{kgephipl}) (Eq. (\ref{kgephim})), we still get the same positively (negatively) charged particle. With this, restricting our analysis to one particle is more straightforward.

Since the corresponding charge density of $\Phi_{+}$ ($\Phi_{-}$) is positive (negative) definite, we can treat $|\phi_{\pm}|^{2}$ as a probability density (again, density in momentum space in our present discussion). That is, from the total charge of the positively charged or negatively charged particle
\begin{align}
\int_{-\infty}^{\infty} \rho_{\pm} dp &= \pm e \nonumber\\
\int_{-\infty}^{\infty} \Phi_{\pm}^{\dagger}\sigma_{3}\Phi_{\pm}dp &= \pm 1\nonumber\\
\int_{-\infty}^{\infty} |\phi_{\pm}|^{2}dp &=  1\nonumber
\end{align}
we see that $|\phi_{\pm}|^{2}$ is normalized to unity. Also, the second line can be interpreted as the normalization of the state $\Phi_{\pm}$, or $<\Phi_{\pm}|\Phi_{\pm}>_{\Phi} = \pm 1$. This implies that we have an inner product defined as $<\Phi_{1}|\Phi_{2}>_{\Phi} = \int_{-\infty}^{\infty} \Phi_{1}^{\dagger}\sigma_{3}\Phi_{2}dp$ for some $\Phi_{1}$ and $\Phi_{2}$ in the (implied) system Hilbert space of states $\mathcal{H}$
\begin{equation}\label{hilbspace}
\mathcal{H} = \mathcal{L}^{2}(\mathbb{R})\otimes\mathbb{C}^{2} : \left\{ \Phi \in \mathcal{H} \left| \int_{-\infty}^{\infty} \Phi^{\dagger}\sigma_{3}\Phi dp < \infty \right. \right\}
\end{equation}
Restricting our analysis to one particle then is much like splitting the Hilbert space into the positive and negative states and working in one of the subspaces. That is, we have $\mathcal{H}=\mathcal{H}_{+}\oplus\mathcal{H}_{-}$ where the elements of $\mathcal{H}_{+}$ ($\mathcal{H}_{-}$), in the $\Phi$-representation, all have the form $\tilde{\Phi}_{+} = (\tilde{\phi}_{+}\; 0)^{T}$ ($\tilde{\Phi}_{-} = (0\; \tilde{\phi}_{-})^{T}$) so that if we are studying a positively (negatively) charged particle, then we only work on the Hilbert (sub)space $\mathcal{H}_{+}$ ($\mathcal{H}_{-}$). 

We may also consider operators in this Hilbert space $\mathcal{H}$. In general, an operator $\hat{A}$ may be decomposed as a sum of a diagonal operator and a non-diagonal operator $\hat{A} = [\hat{A}] + \{\hat{A}\}$ where, $[\hat{A}]$ is the diagonal operator (so-called even operator) and $\{\hat{A}\}$ is the non-diagonal operator (so-called odd operator) \cite{art,book}. If we are only considering one particle, then true one-particle operators should not map any element of $\mathcal{H}_{+}$ into $\mathcal{H}_{-}$ and vice-versa so that the charge sign of the particle is preserved. In the $\Phi$-representation, it is easily seen that these operators are the diagonal operators so that the true one-particle operators are the even operator parts. That is, given a general operator $\hat{A}$ in the $\Phi$-representation, its even part $[\hat{A}]$ is the true one particle operator. An example of an even operator is the Hamiltonian $\mathcal{\hat{H}}_{\Phi}$ and the momentum operator. The position operator, however, is not and we need to take its even part if we wish to use it in one particle analysis.

\section{Constructing $\mathcal{\hat{T}}_{\Psi}$}
\label{constr}

In this section, we attempt to construct a Time of Arrival operator canonically conjugate with the system Hamiltonian with the correct non-relativistic limit. For simplicity, we begin with the canonical commutation relation
\begin{equation}\label{ccr}
\left[\mathcal{\hat{H}}_{\Psi}, \mathcal{\hat{T}}_{\Psi}\right]\Psi = i\hbar \Psi
\end{equation}
where, $\mathcal{\hat{H}}_{\Psi}$ is still the Hamiltonian in Eq. (\ref{kgesch}) in representation-less form. It is explicitly given by
\begin{align}\label{ham}
\mathcal{\hat{H}}_{\Psi} &= 
\begin{pmatrix}
 1 & 1\\
-1 & -1 \\
\end{pmatrix}
\frac{\hat{p}^{2}}{2m_{0}} + 
\begin{pmatrix}
 1 & 0\\
 0 & -1 \\
\end{pmatrix}
m_{0}c^{2}\hat{I} \nonumber\\
&= (\sigma_{3}+i\sigma_{2})\frac{\hat{p}^{2}}{2m_{0}} + \sigma_{3}m_{0}c^{2}, \nonumber\\
\end{align}
the $\sigma_{j}$'s are given by Eq. (\ref{taus}), and $\Psi$ is a two-element column vector. We intend to solve for the operator $\mathcal{\hat{T}}_{\Psi}$ by letting
\begin{equation}\label{texp}
\mathcal{\hat{T}}_{\Psi} = \sum_{m,n} A_{m,n} \hat{T}_{m,n}
\end{equation}
where the $A_{m,n}$'s are unknown but constant $2 \times 2$ matrices, and the $\hat{T}_{m,n}$'s are explicitly given by
\begin{align}\label{tmndef}
\hat{T}_{m,n} &= \frac{1}{2^{n}} \sum_{k=0}^{n} \frac{n!}{k!(n-k)!} \hat{q}^{k}\hat{p}^{m}\hat{q}^{n-k} \nonumber\\
&= \frac{1}{2^{m}} \sum_{j=0}^{m} \frac{m!}{j!(m-j)!} \hat{p}^{j}\hat{q}^{n}\hat{p}^{m-j} \nonumber\\
\end{align}
which can be extended for either negative $m$ or $n$ \cite{ben1, ben2}. The $\hat{T}_{m,n}$'s form a complete linearly independent set and are just the Weyl-ordered quantizations of the one dimensional monomials of the classical position and momenta $q^{n}p^{m}$. In a recent paper \cite{tmndom}, it was shown that the $\hat{T}_{-m,n}$'s for positive $m$ and $n$ are densely-defined operators in the Hilbert space $\mathcal{L}^{2}(\mathbb{R})$ and are thus, meaningful quantum mechanical operators. We use the first equality in Eq (\ref{tmndef}) as our expression for the $\hat{T}_{m,n}$'s and to simplify our calculations, we let the upper limit of the sum over $k$ be positive infinity since the binomial coefficient $\frac{n!}{k!(n-k)!}$ vanishes for $k>n$.
Since the $\sigma_{j}$'s are linearly independent and form a complete set, we can write any $2 \times 2$ matrix in terms of them. Specifically, we can write
\begin{equation}\label{amn}
A_{m,n} = \sum_{j=0}^{3}\alpha_{j}^{m,n}\sigma_{j}
\end{equation}
where the $\alpha_{j}^{m,n}$'s are just unknown scalars. 

Substituting Eq (\ref{texp}) and Eq (\ref{amn}) into the left hand side of Eq (\ref{ccr}), we have
\begin{align}
\left[\mathcal{\hat{H}}_{\Psi}, \mathcal{\hat{T}}_{\Psi}\right]\Psi &= \sum_{m,n}\left(\left[(\sigma_{3}+i\sigma_{2})\frac{\hat{p}^{2}}{2m_{0}}, A_{m,n} \hat{T}_{m,n}\right] + \left[\sigma_{3}m_{0}c^{2}, A_{m,n} \hat{T}_{m,n}\right] \right)\Psi \nonumber \\
&= \sum_{m,n} \left( \frac{1}{2m_{0}}\left( (\sigma_{3}+i\sigma_{2}) A_{m,n} \hat{p}^{2}\hat{T}_{m,n} - A_{m,n}(\sigma_{3}+i\sigma_{2})\hat{T}_{m,n}\hat{p}^{2} \right)\right. \nonumber\\
& \;\;\;\;\;\;\;\;\;\;\;\;  \left. + m_{0}c^{2}\left[\sigma_{3}, A_{m,n} \right]\hat{T}_{m,n} \right)\Psi \nonumber\\
&= \sum_{m,n} \left( \frac{(\sigma_{3}+i\sigma_{2}) A_{m,n}}{2m_{0}} \left( \hat{T}_{m+2,n} - i\hbar n \hat{T}_{m+1,n-1} - \frac{\hbar^{2}}{4} n(n-1)\hat{T}_{m,n-2} \right) \right.\nonumber\\
& \;\; - \frac{A_{m,n} (\sigma_{3}+i\sigma_{2}) }{2m_{0}} \left( \hat{T}_{m+2,n} + i\hbar n \hat{T}_{m+1,n-1} - \frac{\hbar^{2}}{4} n(n-1)\hat{T}_{m,n-2} \right) \nonumber\\
& \;\; \left. + m_{0}c^{2}\left[\sigma_{3}, A_{m,n} \right]\hat{T}_{m,n} \right)\Psi \nonumber\\
&= \sum_{m,n} \left( \left[\sigma_{3}+i\sigma_{2}, \frac{1}{2m_{0}}A_{m-2,n}-\frac{\hbar^{2}(n+2)(n+1)}{8m_{0}}A_{m,n+2} \right] \right. \nonumber\\
& \;\;  \left. -\frac{i\hbar (n+1)}{2m_0} \left\{\sigma_{3}+i\sigma_{2}, A_{m-1,n+1}\right\}  + m_{0}c^{2}\left[\sigma_{3}, A_{m,n} \right]\right)\hat{T}_{m,n}\Psi \nonumber\\ \nonumber
&= \sum_{m,n} \left( \left(\frac{1}{2m_{0}}\alpha_{1}^{m-2,n}-\frac{\hbar^{2}(n+2)(n+1)}{8m_{0}}\alpha_{1}^{m,n+2} \right)2(\sigma_{3}+i\sigma_{2}) \right.\nonumber\\
& \;\; -\left(\frac{1}{2m_{0}}\alpha_{2}^{m-2,n}-\frac{\hbar^{2}(n+2)(n+1)}{8m_{0}}\alpha_{2}^{m,n+2} \right)2i\sigma_{1}\nonumber\\
& \;\; -\left(\frac{1}{2m_{0}}\alpha_{3}^{m-2,n}-\frac{\hbar^{2}(n+2)(n+1)}{8m_{0}}\alpha_{3}^{m,n+2} \right)2\sigma_{1}\nonumber\\
& \;\; -\frac{i\hbar (n+1)}{2m_0}\left( \alpha_{0}^{m-1,n+1}2(\sigma_{3}+i\sigma_{2}) + \alpha_{2}^{m-1,n+1}2i\sigma_{0} + \alpha_{3}^{m-1,n+1}2\sigma_{0} \right)\nonumber\\
& \;\; \left. + m_{0}c^{2}\left( \alpha_{1}^{m,n}2i\sigma_{2} - \alpha_{2}^{m,n}2i\sigma_{1} \right) \right)\hat{T}_{m,n}\Psi\nonumber\\
&= \sum_{m,n} i\hbar \sigma_{0} \delta_{m,0}\delta_{n,0}\hat{T}_{m,n}\Psi\nonumber\\
\end{align}
where the fourth equality is obtained by shifting indices and some rearrangement of terms, while the last equality is the right hand side of Eq (\ref{ccr}). Collecting the coefficients of the $\sigma_{j}$'s, we arrive at four equations for the $\alpha_{j}^{m,n}$'s
\begin{eqnarray}
&& -\frac{i\hbar (n+1)}{2m_0}\left(\alpha_{2}^{m-1,n+1}2i + \alpha_{3}^{m-1,n+1}2 \right) = i\hbar \delta_{m,0}\delta_{n,0} \label{a1} \\
&& -2\left(\frac{\alpha_{3}^{m-2,n} + i \alpha_{2}^{m-2,n}}{2m_{0}}-\frac{\hbar^{2}(n+2)(n+1)}{8m_{0}} \right. \nonumber\\
&&\;\;\;\;\;\;\;\;\;\;\;\;\; \left. \times \left( \alpha_{3}^{m,n+2} + i \alpha_{2}^{m,n+2} \right)\right) - 2im_{0}c^{2}\alpha_{2}^{m,n} = 0\label{a2}\\
&& 2i\left(\frac{1}{2m_{0}} \alpha_{1}^{m-2,n} - \frac{\hbar^{2}(n+2)(n+1)}{8m_{0}} \alpha_{1}^{m,n+2} \right) \nonumber\\
&&\;\;\;\;\;\;\;\;\;\;\;\;\; -2i\frac{i\hbar (n+1)}{2m_0} \alpha_{0}^{m-1,n+1} + 2i m_{0}c^{2} \alpha_{1}^{m,n} = 0 \label{a3}\\
&& 2\left(\frac{1}{2m_{0}} \alpha_{1}^{m-2,n} - \frac{\hbar^{2}(n+2)(n+1)}{8m_{0}} \alpha_{1}^{m,n+2} \right) \nonumber\\ 
&&\;\;\;\;\;\;\;\;\;\;\;\;\; -2\frac{i\hbar (n+1)}{2m_0} \alpha_{0}^{m-1,n+1} = 0 \label{a4} \\ \nonumber
\end{eqnarray}
Combining Eqs (\ref{a3}) and (\ref{a4}) yields $\alpha_{1}^{m,n} = 0$ for any $m,n$ which in turn gives $\alpha_{0}^{m,n \neq 0} = 0$ for any $m$ from Eq (\ref{a4}). Simplifying Eq (\ref{a1}), we get $n\left( \alpha_{3}^{m,n}+i\alpha_{2}^{m,n} \right) = -m_{0}\delta_{m,-1}\delta_{n,1}$ then plugging it in Eq (\ref{a2}):
\begin{displaymath}
\frac{-m_{0}\delta_{m,1}\delta_{n,1}}{2m_0} + \frac{\hbar^{2}n(n+1) m_{0} \delta_{m,-1}\delta_{n,-1}}{8m_0} + im_{0}c^{2}n\alpha_{2}^{m,n} = 0
\end{displaymath}
which gives $\alpha_{2}^{m,n \neq 0} = (2im_{0}c^{2}n)^{-1}\delta_{m,1}\delta_{n,1}$ and then using Eq (\ref{a1}) $\alpha_{3}^{m,n \neq 0} = -(2m_{0}c^{2}n)^{-1}\delta_{m,1}\delta_{n,1} - m_{0}\delta_{m,-1}\delta_{n,1}/n$ for any $m$. Note that we have some arbitrary constants which may not necessarily vanish i.e. $\alpha_{j=0,2,3}^{m,0} \neq 0$ for any $m$. For simplicity however, we wish to take $\mathcal{\hat{T}}_{\Psi}$ as the minimal solution of Eq (\ref{ccr}). That is, we set as many $\alpha_{j}^{m,n}$'s as possible to vanish \cite{ben2} so that the only non-vanishing constants are explicitly given by $\alpha_{2}^{1,1} = (2im_{0}c^{2})^{-1}$, $\alpha_{3}^{1,1} = -(2m_{0}c^{2})^{-1}$, and $\alpha_{3}^{-1,1} = -m_{0}$.


The minimal solution of Eq (\ref{ccr}) is then
\begin{align}
\mathcal{\hat{T}}_{\Psi} &= \left( \alpha_{2}^{1,1}\sigma_{2} + \alpha_{3}^{1,1}\sigma_{3} \right)\hat{T}_{1,1} + \alpha_{3}^{-1,1}\sigma_{3}\hat{T}_{-1,1} \nonumber\\
&= -\frac{1}{2m_{0}c^{2}}(\sigma_{3}+i\sigma_{2})\hat{T}_{1,1} -m_{0}\sigma_{3}\hat{T}_{-1,1}\nonumber\\
&=-\frac{1}{2m_{0}c^{2}}
\begin{pmatrix}
 1 & 1\\
 -1 & -1 \\
\end{pmatrix}
\left( \frac{\hat{p}\hat{q} + \hat{q}\hat{p}}{2} \right) -m_{0}
\begin{pmatrix}
 1 & 0\\
 0 & -1 \\
\end{pmatrix}
\left( \frac{\hat{p}^{-1}\hat{q} + \hat{q}\hat{p}^{-1}}{2} \right)
\end{align}
so that in the Schrodinger-momentum representation ($\Psi-p$ representation), we have the action
\begin{eqnarray}\label{tpsip}
&&\left(\mathcal{\hat{T}}_{\Psi}\Psi\right)(p) = -\frac{1}{2m_{0}c^{2}}
\begin{pmatrix}
 1 & 1\\
 -1 & -1 \\
\end{pmatrix}
\frac{i\hbar}{2}\left( p\frac{\partial}{\partial p}\Psi (p) + \frac{\partial}{\partial p}\left(p\Psi (p)\right)\right) \nonumber\\
&& \;\;\;\;\;\;\;\;\;\;\;\; -m_{0}
\begin{pmatrix}
 1 & 0\\
 0 & -1 \\
\end{pmatrix}
\frac{i\hbar}{2}\left( \frac{1}{p}\frac{\partial}{\partial p}\Psi (p) + \frac{\partial}{\partial p}\left(\frac{1}{p}\Psi (p)\right)\right)\nonumber\\
\end{eqnarray}

\section{Calculation of the even part of $\mathcal{\hat{T}}_{\Phi}$}
\label{onep}
We can calculate the $\Phi -p$ representation of Eq (\ref{tpsip}) by using the transform $\mathcal{\hat{T}}_{\Phi}\Phi = U \mathcal{\hat{T}}_{\Psi} U^{-1}\Phi$ for some trial row vector $\Phi$ where, $U$ and $U^{-1}$ are given by Eq. (\ref{u}).

Consider first $\hat{T}_{1,1}U^{-1}\Phi$ in the momentum representation
\begin{align}
\hat{T}_{1,1}U^{-1}\Phi &= \frac{i\hbar}{2}\left( p\frac{\partial}{\partial p}\left(U^{-1}\Phi\right) + \frac{\partial}{\partial p}\left(p U^{-1}\Phi\right)\right) \nonumber\\
&= \frac{i\hbar}{2}\left( p\left(\frac{\partial }{\partial p} U^{-1} \right)\Phi + p U^{-1}\frac{\partial}{\partial p}\Phi\right. \nonumber\\
& \left. \;\;\;\;\;\; + \left(\frac{\partial }{\partial p} U^{-1} \right)p\Phi +  U^{-1}\frac{\partial}{\partial p}\left(p\Phi\right) \right) \nonumber\\
&= U^{-1}\hat{T}_{1,1}\Phi + i\hbar p\left(-U^{-1}\frac{pc^{2} \sigma_{1}}{2 E_{p}^{2}}\right)\Phi \nonumber\\
&= U^{-1}\left(\hat{T}_{1,1}\Phi- \frac{i\hbar p^{2} c^{2}}{2 E_{p}^{2}}\sigma_{1}\Phi \right) \nonumber\\
\end{align}
and similarly, $\hat{T}_{-1,1}U^{-1}\Phi$
\begin{align}
\hat{T}_{-1,1}U^{-1}\Phi &= \frac{i\hbar}{2}\left( \frac{1}{p}\frac{\partial}{\partial p}\left(U^{-1}\Phi\right) + \frac{\partial}{\partial p}\left(\frac{1}{p} U^{-1}\Phi\right)\right) \nonumber\\
&= \frac{i\hbar}{2}\left( \frac{1}{p}\left(\frac{\partial }{\partial p} U^{-1} \right)\Phi + \frac{1}{p} U^{-1}\frac{\partial}{\partial p}\Phi\right. \nonumber\\
& \left. \;\;\;\;\;\; + \left(\frac{\partial }{\partial p} U^{-1} \right)\frac{1}{p}\Phi +  U^{-1}\frac{\partial}{\partial p}\left(\frac{1}{p}\Phi\right) \right) \nonumber\\
&= U^{-1}\hat{T}_{-1,1}\Phi + i\hbar \frac{1}{p}\left(-U^{-1}\frac{pc^{2} \sigma_{1}}{2 E_{p}^{2}}\right)\Phi \nonumber\\
&= U^{-1}\left(\hat{T}_{-1,1}\Phi- \frac{i\hbar c^{2}}{2 E_{p}^{2}}\sigma_{1}\Phi \right) \nonumber\\
\end{align}
So that we can calculate:
\begin{align}
\mathcal{\hat{T}}_{\Phi}\Phi &= U \mathcal{\hat{T}}_{\Psi} U^{-1}\Phi \nonumber\\
&= -\frac{1}{2m_{0}c^{2}}U(\sigma_{3}+i\sigma_{2})\hat{T}_{1,1}U^{-1}\Phi -m_{0}U\sigma_{3}\hat{T}_{-1,1}U^{-1}\Phi \nonumber\\
&= -\frac{1}{2m_{0}c^{2}}U(\sigma_{3}+i\sigma_{2})U^{-1}\left(\hat{T}_{1,1}\Phi- \frac{i\hbar p^{2} c^{2}}{2 E_{p}^{2}}\sigma_{1}\Phi \right)\nonumber\\
& \;\;\;\;\;\;\;\;\; -m_{0}U\sigma_{3}U^{-1}\left(\hat{T}_{-1,1}\Phi- \frac{i\hbar c^{2}}{2 E_{p}^{2}}\sigma_{1}\Phi \right) \nonumber\\
&= -\frac{1}{2m_{0}c^{2}}\left(\frac{m_{0}c^{2}}{E_{p}}(\sigma_{3}+i\sigma_{2})\right)\left(\hat{T}_{1,1}\Phi- \frac{i\hbar p^{2} c^{2}}{2 E_{p}^{2}}\sigma_{1}\Phi \right)\nonumber\\
& \;\; -m_{0}\left(\frac{m_{0}c^{2}}{E_{p}}\sigma_{3} + \frac{p^{2}}{2m_{0}E_{p}}(\sigma_{3}-i\sigma_{2})\right)\left(\hat{T}_{-1,1}\Phi- \frac{i\hbar c^{2}}{2 E_{p}^{2}}\sigma_{1}\Phi \right) \nonumber\\
&= \sigma_{3}\left[ -\frac{1}{2E_{p}}\hat{T}_{1,1} - \left(\frac{m_{0}^{2}c^{2}}{E_{p}}+\frac{p^{2}}{2E_{p}} \right)\hat{T}_{-1,1} \right]\Phi \nonumber\\
& \;\;\;\;\;\;\;\;\; + \sigma_{2}\left[-\frac{i}{2E_{p}}\hat{T}_{1,1} + \frac{i p^{2}}{2E_{p}}\hat{T}_{-1,1} - \frac{\hbar}{2E_{p}} \right]\Phi\nonumber\\
\end{align}

The even part of $\mathcal{\hat{T}}_{\Phi}$, denoted by $[\mathcal{\hat{T}}_{\Phi}]$, is the diagonal part of $\mathcal{\hat{T}}_{\Phi}$. $[\mathcal{\hat{T}}_{\Phi}]$ represents the operator part that does not mix the positive and negative states and is therefore a true one-particle operator. Explicitly in $\Phi - p$ representation, its action is given by:
\begin{eqnarray}\label{evenT}
&&\left(\left[\mathcal{\hat{T}}_{\Phi}\right]\Phi\right)(p) = - \frac{1}{2E_{p}}
\begin{pmatrix}
 1 & 0\\
 0 & -1 \\
\end{pmatrix}
\frac{i\hbar}{2}\left( p\frac{\partial}{\partial p}\Phi (p) + \frac{\partial}{\partial p}\left(p\Phi (p)\right)\right) \nonumber\\
&& \;\;\;\;\;\;\;\;\;\;\;\; -\left(\frac{m_{0}^{2} c^{2}}{2 E_{p}} + \frac{E_{p}}{2c^{2}}\right)
\begin{pmatrix}
 1 & 0\\
 0 & -1 \\
\end{pmatrix}
\frac{i\hbar}{2}\left( \frac{1}{p}\frac{\partial}{\partial p}\Phi (p) + \frac{\partial}{\partial p}\left(\frac{1}{p}\Phi (p)\right)\right)\nonumber\\
\end{eqnarray}
Being a one-particle operator, we can interpret $\left[\mathcal{\hat{T}}_{\Phi}\right] \equiv \mathcal{\hat{T}}$ as two independent operators, one for positively charged states and another for negatively charged states. Considering the operator for positive states and taking the non-relativistic limit ($c\rightarrow \infty$), we have $E_{p}\rightarrow \infty$ and $E_{p}/c^{2} \rightarrow m_{0}$ so that the operator reduces to $-m_{0}\hat{T}_{-1,1}$. This is the non-relativistic operator for the free TOA at the origin (c.f. \cite{abohm, gal4}). Moreover, it can be shown that $\mathcal{\hat{T}}$ is still canonically conjugate to $[\mathcal{\hat{H}}_{\Phi}] = \mathcal{\hat{H}}_{\Phi} = E_{p}\sigma_{3}$. That is, we still have $[\mathcal{\hat{H}}_{\Phi}, \mathcal{\hat{T}}] \Phi = i\hbar\Phi$ for an arbitrary $\Phi$.


\section{$\mathcal{\hat{T}}$ is a symmetric operator}
\label{symmop}
For ease of notation, we write $ \hat{T} \equiv \sigma_{3} \mathcal{\hat{T}} $ . If $\mathcal{\hat{T}}$ is symmetric (or Hermitian in the sense of \cite{book}), we should have $\left<\Phi_{a}\left|\mathcal{\hat{T}}\Phi_{b}\right.\right>_{\Phi} = \left<\left.\mathcal{\hat{T}}\Phi_{a}\right|\Phi_{b}\right>_{\Phi}$ for some $\Phi_{a} = (\phi_{a1}\;\phi_{a2})^{T}$ and $\Phi_{b} = (\phi_{b1}\;\phi_{b2})^{T}$. For simplicity, we let the $\phi$'s vanish at infinity so that $\Phi_{a}$ and $\Phi_{b}$ are in the Hilbert space Eq. (\ref{hilbspace}). We first calculate:
\begin{align}
p\frac{\partial }{\partial p}\frac{1}{E_{p}} &= -p\frac{1}{E_{p}^{2}}\frac{\partial }{\partial p}E_{p} = -\frac{p^{2}c^{2}}{E_{p}^{3}}\nonumber\\
\frac{1}{p}\frac{\partial }{\partial p}\left(\frac{m_{0}^{2} c^{2}}{ E_{p}} + \frac{E_{p}}{c^{2}}\right) &= \frac{1}{p}\left(-\frac{pc^{2}m_{0}^{2}c^{2}}{E_{p}^{3}}+\frac{pc^{2}}{E_{p}c^{2}} \right) = \frac{-m_{0}^{2}c^{4}+E_{p}^{2}}{E_{p}^{3}} = \frac{p^{2}c^{2}}{E_{p}^{3}}\nonumber\\
\nonumber
\end{align}
Then consider the integral:
\begin{align}
\int_{-\infty}^{\infty}\phi_{a}^{*}\hat{T}\phi_{b}dp &= -\frac{i\hbar}{4}\int_{-\infty}^{\infty}\phi_{a}^{*}\frac{1}{E_{p}}\left( p\frac{\partial}{\partial p}\phi_{b} + \frac{\partial}{\partial p}\left(p\phi_{b}\right) \right) dp \nonumber\\
&\;\;\;\; -\frac{i\hbar}{4}\int_{-\infty}^{\infty}\phi_{a}^{*}\left(\frac{m_{0}^{2} c^{2}}{ E_{p}} + \frac{E_{p}}{c^{2}}\right)\left(\frac{1}{p}\frac{\partial}{\partial p}\phi_{b}+\frac{\partial}{\partial p}\left(\frac{1}{p}\phi_{b}\right) \right)dp  \nonumber\\
&= \frac{i\hbar}{4}\int_{-\infty}^{\infty}\left(\frac{\partial}{\partial p}\left(\frac{1}{E_{p}}p\phi_{a}^{*}\right)\phi_{b}+ p\frac{\partial}{\partial p}\left(\frac{1}{E_{p}}\phi_{a}^{*}\right)\phi_{b} \right)dp\nonumber\\
&\;\;\;\; + \frac{i\hbar}{4}\int_{-\infty}^{\infty}\left( \frac{\partial}{\partial p}\left[\left(\frac{m_{0}^{2} c^{2}}{ E_{p}} + \frac{E_{p}}{c^{2}}\right)\frac{1}{p}\phi_{a}^{*}\right]\phi_{b} \right. \nonumber\\
&\;\;\;\;\;\;\; \left. + \frac{1}{p}\frac{\partial}{\partial p}\left[ \left(\frac{m_{0}^{2} c^{2}}{ E_{p}} + \frac{E_{p}}{c^{2}}\right)\phi_{a}^{*}\right]\phi_{b}\right)dp \nonumber\\
&= \frac{i\hbar}{4}\int_{-\infty}^{\infty}\frac{1}{E_{p}}\left(\frac{\partial}{\partial p}\left(p\phi_{a}^{*}\right)+ p\frac{\partial}{\partial p}\left(\phi_{a}^{*}\right) \right)\phi_{b} dp\nonumber\\
&\;\;\;\; + \frac{i\hbar}{4}\int_{-\infty}^{\infty}\left(\frac{m_{0}^{2} c^{2}}{ E_{p}} + \frac{E_{p}}{c^{2}}\right)\left(\frac{\partial}{\partial p}\left(\frac{1}{p}\phi_{a}^{*}\right) + \frac{1}{p}\frac{\partial}{\partial p}\left(\phi_{a}^{*}\right) \right)\phi_{b}dp  \nonumber\\
&\;\;\;\; + \frac{i\hbar}{2}\int_{-\infty}^{\infty}\left[ p\phi_{a}^{*}\phi_{b} \frac{\partial }{\partial p}\frac{1}{E_{p}}  + \frac{1}{p}\phi_{a}^{*}\phi_{b}\frac{\partial }{\partial p}\left(\frac{m_{0}^{2} c^{2}}{ E_{p}} + \frac{E_{p}}{c^{2}}\right) \right]dp \nonumber\\
&=\int_{-\infty}^{\infty}\left(\hat{T}\phi_{a}\right)^{*}\phi_{b}dp + \frac{i\hbar}{2}\int_{-\infty}^{\infty}\phi_{a}^{*}\phi_{b} \left[-\frac{p^{2}c^{2}}{E_{p}^{3}} + \frac{p^{2}c^{2}}{E_{p}^{3}}\right]dp\nonumber\\
&=\int_{-\infty}^{\infty}\left(\hat{T}\phi_{a}\right)^{*}\phi_{b}dp \nonumber\\
\end{align}
so that we can calculate (noting that $\sigma_{3}^{\dagger} = \sigma_{3}$)
\begin{align}
\left<\Phi_{a}\left|\mathcal{\hat{T}}\Phi_{b}\right.\right>_{\Phi} &= \int_{-\infty}^{\infty}\Phi_{a}^{\dagger}\sigma_{3}\mathcal{\hat{T}}\Phi_{b} dp \nonumber\\
& = \int_{-\infty}^{\infty}\phi_{a1}^{*}\hat{T}\phi_{b1}dp + \int_{-\infty}^{\infty}\phi_{a2}^{*}\hat{T}\phi_{b2}dp\nonumber\\
& = \int_{-\infty}^{\infty}\left(\hat{T}\phi_{a1}\right)^{*}\phi_{b1}dp + \int_{-\infty}^{\infty}\left(\hat{T}\phi_{a2}\right)^{*}\phi_{b2}dp\nonumber\\
& = \int_{-\infty}^{\infty} 
\begin{pmatrix}
\left(\hat{T}\phi_{a1}\right)^{*}  \left(\hat{T}\phi_{a2}\right)^{*} \\
\end{pmatrix}
\begin{pmatrix}
\phi_{b1} \\
\phi_{b2} \\
\end{pmatrix}
dp\nonumber\\
& = \int_{-\infty}^{\infty}\left(\sigma_{3}\mathcal{\hat{T}}\Phi_{a}\right)^{\dagger}\Phi_{b} dp \nonumber\\
& = \int_{-\infty}^{\infty}\left(\mathcal{\hat{T}}\Phi_{a}\right)^{\dagger}\sigma_{3}\Phi_{b} dp \nonumber\\
& = \left<\left.\mathcal{\hat{T}}\Phi_{a}\right|\Phi_{b}\right>_{\Phi}\nonumber\\
\end{align}
where, $A^{\dagger}$ is just the complex conjugate of the transpose of the matrix $A$. $\mathcal{\hat{T}}$ is then symmetric. This means that its expectation values are real.

\section{The eigenfunctions of $\mathcal{\hat{T}}$}
\label{eigfsec}
Now, we solve for the eigenfunctions of $\mathcal{\hat{T}}$ by considering positive states $\Phi_{+} = (1 \; 0)^{T} \phi_{+}(p)$ and negative states $\Phi_{-} = (0 \; 1)^{T} \phi_{-}(p)$ so that $\mathcal{\hat{T}}\Phi_{\pm} = \tau\Phi_{\pm}$ where $\tau$ is their corresponding eigenvalue. Using Eq (\ref{evenT}), we get
\begin{eqnarray}\label{eigfdiff}
&& \tau \phi_{\pm} = -\frac{1}{2E_{p}}\frac{i\hbar}{2}\left( p\frac{\partial}{\partial p}\left(\pm \phi_{\pm}\right) + \frac{\partial}{\partial p}\left(\pm p\phi_{\pm}\right)\right) \nonumber\\
&& -\left(\frac{m_{0}^{2} c^{2}}{2 E_{p}} + \frac{E_{p}}{2c^{2}}\right)\frac{i\hbar}{2}\left( \frac{1}{p}\frac{\partial}{\partial p}\left(\pm \phi_{\pm}\right) + \frac{\partial}{\partial p}\left(\pm \frac{1}{p}\phi_{\pm}\right)\right) \nonumber\\
&& \pm i\frac{4 E_{p}\tau}{\hbar}p^{2}\phi_{\pm} = 2p^{3} \frac{\partial \phi_{\pm}}{\partial p} + p^{2}\phi_{\pm} + \left(2m_{0}^{2}c^{2}+p^{2}\right)\left(2p\frac{\partial \phi_{\pm}}{\partial p} - \phi_{\pm} \right)\nonumber\\
&& \pm i\frac{2 E_{p}\tau}{\hbar}p^{2}\phi_{\pm} = 2p^{3} \frac{\partial \phi_{\pm}}{\partial p} + 2m_{0}^{2}c^{2}p \frac{\partial \phi_{\pm}}{\partial p} -m_{0}^{2}c^{2}\phi_{\pm}\nonumber\\
&& \left(\pm i\frac{2 E_{p}\tau}{\hbar}p^{2} + m_{0}^{2}c^{2}\right)\phi_{\pm} = \frac{2p E_{p}^{2}}{c^{2}}\frac{\partial \phi_{\pm}}{\partial p}\nonumber\\
&& \frac{\partial \phi_{\pm}}{\partial p} = \left[ \pm i \frac{p c^{2} \tau}{E_{p}\hbar} + \frac{m_{0}^{2}c^{4}}{2p E_{p}^{2}}\right]\phi_{\pm} \nonumber\\
&& \frac{\partial \phi_{\pm}}{\partial p} = \left[ \pm i \frac{c\tau}{\hbar}\frac{p}{\sqrt{p^{2}+m_{0}^{2}c^{2}}} + \frac{m_{0}^{2}c^{2}}{2}\frac{1}{p(p^{2}+m_{0}^{2}c^{2})}\right]\phi_{\pm}\nonumber\\
\end{eqnarray}
We treat the eigenfunction $\phi_{\pm}$ as a distribution and solve Eq (\ref{eigfdiff}) for $p>0$ and $p<0$ (mirroring \cite{muga3}) which gives us
\begin{displaymath}
\tilde{\phi}_{\pm,\tau}^{(p_{\lambda}=\pm 1)} = A \sqrt{\frac{|p|c}{E_{p}}}\exp\left(\pm i\frac{E_{p}\tau}{\hbar}\right)\Theta(p_{\lambda}p)
\end{displaymath}
where, $A$ is some constant and $\Theta(p)$ is the Heaviside step function. We choose our eigenfunctions as the sum and difference of these so that our eigenfunctions would have definite parity in momentum. That is,
\begin{align}\label{eigf}
\phi_{\pm,\tau}^{(+)} &= \tilde{\phi}_{\pm,\tau}^{(+1)} + \tilde{\phi}_{\pm,\tau}^{(-1)} = A \sqrt{\frac{|p|c}{E_{p}}}\exp\left(\pm i\frac{E_{p}\tau}{\hbar}\right)\nonumber\\
\phi_{\pm,\tau}^{(-)} &= \tilde{\phi}_{\pm,\tau}^{(+1)} - \tilde{\phi}_{\pm,\tau}^{(-1)} = A \sqrt{\frac{|p|c}{E_{p}}}\exp\left(\pm i\frac{E_{p}\tau}{\hbar}\right)\mbox{sgn}(p)\nonumber\\
\end{align}
Note that $\mbox{sgn}(p)$ is the sign function. This choice is also made in order to reproduce the dynamical behaviours of the so-called nodal and non-nodal eigenfunctions of the Confined Time of Arrival operators \cite{gal5,gal6} as we will show in section \ref{dynamics}. Explicitly, our eigenfunctions are then
\begin{align}\label{eigft0}
\Phi_{\lambda = \pm 1,\tau}^{(+)}(p) &= 
\begin{pmatrix}
\Theta(\lambda) \\
\Theta(-\lambda) \\
\end{pmatrix}
\phi_{\lambda,\tau}^{(+)}(p)\nonumber\\
&= 
\begin{pmatrix}
\Theta(\lambda) \\
\Theta(-\lambda) \\
\end{pmatrix}
A \sqrt{\frac{|p|c}{E_{p}}}\exp\left(\lambda i\frac{\tau E_{p}}{\hbar}\right)\nonumber\\
\Phi_{\lambda = \pm 1,\tau}^{(-)}(p) &= 
\begin{pmatrix}
\Theta(\lambda) \\
\Theta(-\lambda) \\
\end{pmatrix}
\phi_{\lambda,\tau}^{(-)}(p)\nonumber\\
&= 
\begin{pmatrix}
\Theta(\lambda) \\
\Theta(-\lambda) \\
\end{pmatrix}
A \sqrt{\frac{|p|c}{E_{p}}}\exp\left(\lambda i\frac{\tau E_{p}}{\hbar}\right)\mbox{sgn}(p)\nonumber\\
\end{align}
For a given eigenvalue $\tau$, $\mathcal{\hat{T}}$ then has two degenerate eigenfunctions $\Phi_{\lambda,\tau}^{(+)}(p)$ and $\Phi_{\lambda,\tau}^{(-)}(p)$ describing a particle with charge $\lambda e$.

\section{Completeness of the $\Phi_{\lambda,\tau}^{(n)}(p)$'s}
\label{complete}
If the eigenfunctions $\Phi_{\lambda,\tau}^{(n)}(p)$ given by Eq (\ref{eigft0}) are complete, then we should have the outer product
\begin{equation}\label{compl}
\sum_{\lambda=\pm 1}\sum_{n=\pm}\int_{-\infty}^{\infty}\Phi_{\lambda,\tau}^{(n)}(p)\Phi_{\lambda,\tau}^{(n)\dagger}(p')d\tau = \sigma_{0}\delta(p-p')
\end{equation}
Consider first the integral
\begin{align}
\int_{-\infty}^{\infty}\phi_{\lambda,\tau}^{(+)}(p)\phi_{\lambda,\tau}^{(+)*}(p')d\tau &= |A|^{2}\sqrt{\frac{|pp'|c^{2}}{E_{p}E_{p'}}}\int_{-\infty}^{\infty}e^{\lambda i \tau (E_{p}-E_{p'})/\hbar}d\tau\nonumber\\
&=|A|^{2}\sqrt{\frac{|pp'|c^{2}}{E_{p}E_{p'}}}2\pi\hbar\delta(E_{p}-E_{p'})\nonumber\\
&=2\pi\hbar |A|^{2}\frac{|p|c}{E_{p}} \frac{E_{p}}{|p|c^{2}}\left[\delta(p-p')+\delta(p+p')\right]\nonumber\\
&=\frac{2\pi\hbar}{c} |A|^{2}\left[\delta(p-p')+\delta(p+p')\right]\nonumber\\
\end{align}
and similarly,
\begin{align}
\int_{-\infty}^{\infty}\phi_{\lambda,\tau}^{(-)}(p)\phi_{\lambda,\tau}^{(-)*}(p')d\tau &= |A|^{2}\sqrt{\frac{|pp'|c^{2}}{E_{p}E_{p'}}}\mbox{sgn}(p)\mbox{sgn}(p')\int_{-\infty}^{\infty}e^{\lambda i\tau(E_{p}-E_{p'})/\hbar}d\tau\nonumber\\
&=|A|^{2}\sqrt{\frac{|pp'|c^{2}}{E_{p}E_{p'}}}\mbox{sgn}(p)\mbox{sgn}(p')2\pi\hbar\delta(E_{p}-E_{p'})\nonumber\\
&=2\pi\hbar |A|^{2}\frac{|p|c}{E_{p}}\mbox{sgn}(p)\mbox{sgn}(p') \frac{E_{p}\left[\delta(p-p')+\delta(p+p')\right]}{|p|c^{2}}\nonumber\\
&=2\pi\hbar |A|^{2}\frac{|p|c}{E_{p}} \frac{E_{p}}{|p|c^{2}}\left[\delta(p-p')-\delta(p+p')\right]\nonumber\\
&=\frac{2\pi\hbar}{c} |A|^{2} \left[\delta(p-p')-\delta(p+p')\right]\nonumber\\
\end{align}
where, the second to the last line was obtained since $\delta(p+p')$ implies that $p$ and $p'$ have opposite signs. The left hand side of Eq (\ref{compl}) then reads
\begin{align}
\sum_{\lambda=\pm 1}\sum_{n=\pm}
\begin{pmatrix}
\Theta(\lambda) \\
\Theta(-\lambda) \\
\end{pmatrix}
\begin{pmatrix}
\Theta(\lambda) & \Theta(-\lambda) \\
\end{pmatrix}
&\int_{-\infty}^{\infty}\phi_{\lambda,\tau}^{(n)}(p)\phi_{\lambda,\tau}^{(n)*}(p')d\tau \nonumber\\
&= \sum_{\lambda=\pm 1}
\begin{pmatrix}
\Theta(\lambda) & 0 \\
 0 & \Theta(-\lambda) \\
\end{pmatrix}
\frac{4\pi\hbar}{c} |A|^{2}\delta(p-p')\nonumber\\
&= \frac{4\pi\hbar}{c} |A|^{2}\sigma_{0}\delta(p-p')\nonumber\\
&= \sigma_{0}\delta(p-p')\nonumber\\
\end{align}
where, $\Theta(\lambda)\Theta(-\lambda) = 0$, and we let $A=\sqrt{\frac{c}{4\pi\hbar}}$ so that the eigenfunctions
\begin{align}\label{normeigft}
\Phi_{\lambda,\tau}^{(+)}(p)
&= 
\begin{pmatrix}
\Theta(\lambda) \\
\Theta(-\lambda) \\
\end{pmatrix}
\sqrt{\frac{c}{4\pi\hbar}} \sqrt{\frac{|p|c}{E_{p}}}\exp\left(\lambda i\frac{\tau E_{p}}{\hbar}\right)\nonumber\\
\Phi_{\lambda,\tau}^{(-)}(p) 
&= 
\begin{pmatrix}
\Theta(\lambda) \\
\Theta(-\lambda) \\
\end{pmatrix}
\sqrt{\frac{c}{4\pi\hbar}} \sqrt{\frac{|p|c}{E_{p}}}\exp\left(\lambda i\frac{\tau E_{p}}{\hbar}\right)\mbox{sgn}(p)\nonumber\\
\end{align}
form a complete set. It can be shown that upon time evolution, the set of eigenfunctions Eq (\ref{normeigft}) still form a complete set for any later time $t$.

Before we proceed, let us consider the non-relativistic limit of Eq (\ref{normeigft}) for $p>0$ and $p<0$. That is, as $c\rightarrow \infty$, we have $E_{p} \sim m_{0}c^{2} + \frac{p^{2}}{2m_{0}}$ and $E_{p}/c^{2} \sim m_{0}$ so that 
\begin{displaymath}
\tilde{\Phi}_{\lambda,\tau}^{(p_{\lambda})} \sim \mbox{e}^{\lambda i \frac{\tau m_{0}c^{2}}{\hbar}} \sqrt{\frac{|p|}{m_{0}\hbar}}\exp\left(\lambda i \frac{\tau p^{2} }{2 m_{0}\hbar}\right) \Theta(p_{\lambda}p)
\end{displaymath}
which, apart from a phase factor irrelevant when taking probability densities, are just the eigenfunctions of the free non-relativistic TOA operator $-m_{0}\hat{T}_{-1,1}$ \cite{gal6, galmuga, muga2, muga3}.

\section{The Non-Orthogonality of the $\Phi_{\lambda,\tau}^{(n)}(p)$'s}
\label{nonorthsec}
In this section, we wish to compute for the inner product
\begin{align}\label{nonorth}
\int_{-\infty}^{\infty} \Phi_{\lambda',\tau'}^{(n')\dagger}(p) & \sigma_{3}\Phi_{\lambda,\tau}^{(n)}(p)dp \nonumber\\
&= 
\begin{pmatrix}
\Theta(\lambda') & \Theta(-\lambda') \\
\end{pmatrix}
\begin{pmatrix}
\Theta(\lambda) \\
-\Theta(-\lambda) \\
\end{pmatrix}
\int_{-\infty}^{\infty} \phi_{\lambda',\tau'}^{(n')*}(p)\phi_{\lambda,\tau}^{(n)}(p)dp \nonumber\\
&= \delta_{\lambda,\lambda'}\left( \Theta(\lambda) - \Theta(-\lambda)\right) \int_{-\infty}^{\infty} \phi_{\lambda,\tau'}^{(n')*}(p)\phi_{\lambda,\tau}^{(n)}(p)dp\nonumber\\
&= \delta_{\lambda,\lambda'} \delta_{n,n'}\lambda\int_{-\infty}^{\infty} \phi_{\lambda,\tau'}^{(n)*}(p)\phi_{\lambda,\tau}^{(n)}(p)dp\nonumber\\
\end{align}
Note that when $\lambda \neq \lambda'$ (they differ in signs), $\Theta(\lambda')\Theta(\lambda)$ vanishes so that the product $(\Theta(\lambda')\;\Theta(-\lambda'))(\Theta(\lambda)\;-\Theta(-\lambda))^{T}$ also vanishes. Also note that when $n \neq n'$ the integrand $\phi_{\lambda,\tau'}^{(n')*}(p)\phi_{\lambda,\tau}^{(n)}(p)$ is odd in $p$ so that the integral vanishes. We are then left to compute for the integral in the last equality.
\begin{align}
\int_{-\infty}^{\infty} \phi_{\lambda,\tau'}^{(n)*}(p)\phi_{\lambda,\tau}^{(n)}(p)dp &= \int_{-\infty}^{\infty} \frac{c}{4\pi\hbar}\frac{|p|c}{E_{p}}\mbox{e}^{-\lambda i \frac{\tau'}{\hbar}E_{p}}\mbox{e}^{\lambda i \frac{\tau}{\hbar}E_{p}}dp \nonumber\\
&=\frac{c}{2\pi\hbar}\int_{0}^{\infty}\frac{pc}{E_{p}}\mbox{e}^{\lambda i \frac{(\tau-\tau')}{\hbar}E_{p}}dp\nonumber\\
&= \frac{1}{2\pi\hbar}\int_{m_{0}c^{2}}^{\infty}\exp\left(\lambda i \frac{(\tau-\tau')}{\hbar}E_{p}\right) dE_{p}\nonumber\\
&= \frac{1}{2\pi\hbar}\lim_{\epsilon \rightarrow 0^{+}} \int_{m_{0}c^{2}}^{\infty}\exp\left(-\left(\epsilon - \lambda i \frac{(\tau-\tau')}{\hbar}\right)E_{p}\right) dE_{p}\nonumber\\
&= \frac{1}{2\pi\hbar}\lim_{\epsilon \rightarrow 0^{+}}\frac{\exp\left(-\left(\epsilon - \lambda i \frac{(\tau-\tau')}{\hbar}\right)m_{0}c^{2}\right)}{\left(\epsilon - \lambda i \frac{(\tau-\tau')}{\hbar}\right)}\nonumber\\
&= \frac{1}{2\pi\hbar}\lim_{\epsilon \rightarrow 0^{+}} \mbox{e}^{\lambda i \frac{(\tau-\tau')m_{0}c^{2}}{\hbar}}\frac{\epsilon + \lambda i\frac{(\tau-\tau')}{\hbar} }{\epsilon^{2}+\left(\frac{\tau-\tau'}{\hbar}\right)^{2}}\nonumber\\
\nonumber
\end{align}
where $\exp(-\epsilon)$ was neglected since it is only a multiplicative factor approaching unity. From \cite{deltawolfram}, we recognize a term above as a Dirac Delta.
\begin{align}
\int_{-\infty}^{\infty} \phi_{\lambda,\tau'}^{(n)*}(p)\phi_{\lambda,\tau}^{(n)}(p)dp &= \frac{\mbox{e}^{\lambda i \frac{(\tau-\tau')m_{0}c^{2}}{\hbar}}}{2}\left(\delta(\tau-\tau') + \frac{1}{\pi\hbar}\lim_{\epsilon \rightarrow 0^{+}} \frac{\lambda i \frac{(\tau-\tau')}{\hbar}}{\epsilon^{2}+ \left(\frac{\tau-\tau'}{\hbar}\right)^{2}}\right)\nonumber\\
&= \frac{\delta(\tau-\tau')}{2} + \frac{\lambda i  \mbox{e}^{\lambda i \frac{(\tau-\tau')m_{0}c^{2}}{\hbar}}}{2 \pi (\tau-\tau')}\nonumber\\
\nonumber
\end{align}
Therefore, Eq. (\ref{nonorth}) becomes
\begin{align}
\int_{-\infty}^{\infty} \Phi_{\lambda',\tau'}^{(n')\dagger}(p) & \sigma_{3}\Phi_{\lambda,\tau}^{(n)}(p)dp \nonumber\\
&= \delta_{\lambda,\lambda'} \delta_{n,n'}\lambda\left(\frac{\delta(\tau-\tau')}{2} + \frac{\lambda i  \mbox{e}^{\lambda i \frac{(\tau-\tau')m_{0}c^{2}}{\hbar}}}{2 \pi (\tau-\tau')}\right) \nonumber\\
\end{align}
The eigenfunctions $\Phi_{\lambda,\tau}^{(n)}(p)$ are then non-orthogonal. It can also be shown that upon time evolution, the set of eigenfunctions remain non-orthogonal at any later time $t$. This suggests that the operator $\mathcal{\hat{T}}$ is non-self-adjoint. $\mathcal{\hat{T}}$ is then a maximally symmetric operator. A property which is also true for the non-relativistic case $-m_{0}\hat{T}_{-1,1}$ \cite{gal5, gal6, galmuga, muga1, muga2, muga3}.


\section{The eigenfunctions in the $\Phi - x$ representation and their dynamic behaviour}
\label{dynamics}
Now, we consider the time evolution of Eq (\ref{normeigft}) by solving the Klein-Gordon equation
\begin{equation}\label{kg}
i\hbar\frac{\partial}{\partial t}\Phi_{\lambda,\tau}^{(n)}(p,t) =  \mathcal{\hat{H}}_{\Phi}\Phi_{\lambda,\tau}^{(n)}(p,t) = E_{p}\sigma_{3}\Phi_{\lambda,\tau}^{(n)}(p,t)
\end{equation}
for $\Phi_{\lambda,\tau}^{(n)}(p,t) = (\Theta(\lambda)\; \Theta(-\lambda))^{T}\phi_{\lambda,\tau}^{(n)}(p,t)$ where $\mathcal{\hat{H}}_{\Phi}$ is the $\Phi -p$ representation of the Hamiltonian in Eq (\ref{ham}) and $\Phi_{\lambda,\tau}^{(n)}(p,0)$ is given by Eq (\ref{normeigft}). The solution of Eq (\ref{kg}) is readily seen as
\begin{align}
\phi_{\pm,\tau}^{(+)}(p,t) &= \phi_{\pm,\tau}^{(+)}(p, 0)\exp\left(\mp i\frac{E_{p} t}{\hbar} \right) = \sqrt{\frac{c}{4\pi\hbar}} \sqrt{\frac{|p|c}{E_{p}}}\exp\left(\mp i\frac{(t-\tau)E_{p}}{\hbar}\right)\nonumber\\
\phi_{\pm,\tau}^{(-)}(p,t) &= \phi_{\pm,\tau}^{(-)}(p, 0)\exp\left(\mp i\frac{E_{p} t}{\hbar} \right) = \sqrt{\frac{c}{4\pi\hbar}} \sqrt{\frac{|p|c}{E_{p}}}\exp\left(\mp i\frac{(t-\tau)E_{p}}{\hbar}\right)\mbox{sgn}(p)\nonumber\\
\end{align}
We then write the time-evolved eigenfunctions as
\begin{align}\label{eigft}
\Phi_{\lambda,\tau}^{(+)}(p,t) &= 
\begin{pmatrix}
\Theta(\lambda) \\
\Theta(-\lambda) \\
\end{pmatrix}
\phi_{\lambda,\tau}^{(+)}(p, t)\nonumber\\
&= 
\begin{pmatrix}
\Theta(\lambda) \\
\Theta(-\lambda) \\
\end{pmatrix}
\sqrt{\frac{c}{4\pi\hbar}} \sqrt{\frac{|p|c}{E_{p}}}\exp\left(-\lambda i\frac{(t-\tau)E_{p}}{\hbar}\right)\nonumber\\
\Phi_{\lambda,\tau}^{(-)}(p,t) &= 
\begin{pmatrix}
\Theta(\lambda) \\
\Theta(-\lambda) \\
\end{pmatrix}
\phi_{\lambda,\tau}^{(-)}(p, t)\nonumber\\
&= 
\begin{pmatrix}
\Theta(\lambda) \\
\Theta(-\lambda) \\
\end{pmatrix}
\sqrt{\frac{c}{4\pi\hbar}} \sqrt{\frac{|p|c}{E_{p}}}\exp\left(-\lambda i\frac{(t-\tau)E_{p}}{\hbar}\right)\mbox{sgn}(p)\nonumber\\
\end{align}
We can also rewrite the $\Phi_{\lambda, \tau}^{(n)}(p,t)$'s by letting $q=\frac{p}{m_{0}c}$
\begin{align}\label{eigftq}
\Phi_{\lambda, \tau}^{(+)}(q,t) &= 
\begin{pmatrix}
\Theta(\lambda) \\
\Theta(-\lambda) \\
\end{pmatrix}
\sqrt{\frac{c}{4\pi\hbar}} \sqrt{\frac{|q|}{\sqrt{1+q^2}}}\exp\left(-\lambda i\frac{m_{0}c^{2}}{\hbar}(t-\tau)\sqrt{1+q^{2}}\right)\nonumber\\
\Phi_{\lambda, \tau}^{(-)}(q,t) &= 
\begin{pmatrix}
\Theta(\lambda) \\
\Theta(-\lambda) \\
\end{pmatrix}
\sqrt{\frac{c}{4\pi\hbar}} \sqrt{\frac{|q|}{\sqrt{1+q^2}}}\exp\left(-\lambda i\frac{m_{0}c^{2}}{\hbar}(t-\tau)\sqrt{1+q^{2}}\right)\mbox{sgn}(q)\nonumber\\
\end{align}
We then take its Fourier Transform to obtain the eigenfunctions in the position representation. That is,
\begin{align}
\Phi_{\lambda,\tau}^{(n)}(x,t) &= \frac{1}{\sqrt{2\pi\hbar}}\int_{-\infty}^{\infty}\exp\left(i\frac{px}{\hbar}\right)\Phi_{\lambda,\tau}^{(n)}(p,t)dp\nonumber\\
&=\frac{m_{0}c}{\sqrt{2\pi\hbar}}\int_{-\infty}^{\infty}\exp\left(i\frac{m_{0}c}{\hbar}xq\right)\Phi_{\lambda,\tau}^{(n)}(q,t)dq\nonumber\\
\nonumber
\end{align}
Explicitly, we have
\begin{align}\label{ppxt}
\Phi_{\lambda,\tau}^{(+)}(x,t) &= \frac{m_{0}c}{\sqrt{2\pi\hbar}}\int_{-\infty}^{\infty}\exp\left(i\frac{m_{0}c x}{\hbar}q\right)\Phi_{\lambda,\tau}^{(+)}(q,t)dq\nonumber\\
&=
\begin{pmatrix}
\Theta(\lambda) \\
\Theta(-\lambda) \\
\end{pmatrix}
\frac{m_{0}c}{2\pi\hbar}\sqrt{\frac{c}{2}} \int_{-\infty}^{\infty}\frac{\sqrt{|q|}}{\sqrt[4]{1+q^2}}\mbox{e}^{i\frac{m_{0}c}{\hbar}xq-\lambda i\frac{m_{0}c^{2}}{\hbar}(t-\tau)\sqrt{1+q^{2}}}dq\nonumber\\
&=
\begin{pmatrix}
\Theta(\lambda) \\
\Theta(-\lambda) \\
\end{pmatrix}
\frac{m_{0}c}{\pi\hbar}\sqrt{\frac{c}{2}} \int_{0}^{\infty}\frac{\sqrt{q}}{\sqrt[4]{1+q^2}}\cos\left(\frac{m_{0}c}{\hbar}xq\right)\mbox{e}^{-\lambda i\frac{m_{0}c^{2}}{\hbar}(t-\tau)\sqrt{1+q^{2}}}dq\nonumber\\
&=
\begin{pmatrix}
\Theta(\lambda) \\
\Theta(-\lambda) \\
\end{pmatrix}
\frac{m_{0}c}{\pi\hbar}\sqrt{\frac{c}{2}} \int_{0}^{\infty}\frac{\sqrt{q}}{\sqrt[4]{1+q^2}}\cos\left(\frac{m_{0}c}{\hbar}xq\right)\cos\left( \frac{m_{0}c^{2}}{\hbar}(t-\tau)\sqrt{1+q^{2}}\right)dq\nonumber\\
& -i\lambda
\begin{pmatrix}
\Theta(\lambda) \\
\Theta(-\lambda) \\
\end{pmatrix}
\frac{m_{0}c}{\pi\hbar}\sqrt{\frac{c}{2}} \int_{0}^{\infty}\frac{\sqrt{q}}{\sqrt[4]{1+q^2}}\cos\left(\frac{m_{0}c}{\hbar}xq\right)\sin\left( \frac{m_{0}c^{2}}{\hbar}(t-\tau)\sqrt{1+q^{2}}\right)dq\nonumber\\
&\equiv
\begin{pmatrix}
\Theta(\lambda) \\
\Theta(-\lambda) \\
\end{pmatrix}
\frac{m_{0}c}{\pi\hbar}\sqrt{\frac{c}{2}} (f_{1,\tau}^{(+)}(x,t)-i\lambda f_{2,\tau}^{(+)}(x,t))\nonumber\\
\end{align}
\begin{align}\label{pmxt}
\Phi_{\lambda,\tau}^{(-)}(x,t) &= \frac{m_{0}c}{\sqrt{2\pi\hbar}}\int_{-\infty}^{\infty}\exp\left(i\frac{m_{0}c x}{\hbar}q\right)\Phi_{\lambda,\tau}^{(-)}(q,t)dq\nonumber\\
&=
\begin{pmatrix}
\Theta(\lambda) \\
\Theta(-\lambda) \\
\end{pmatrix}
\frac{m_{0}c}{2\pi\hbar}\sqrt{\frac{c}{2}} \int_{-\infty}^{\infty}\frac{\sqrt{|q|}}{\sqrt[4]{1+q^2}}\mbox{e}^{i\frac{m_{0}c}{\hbar}xq-\lambda i\frac{m_{0}c^{2}}{\hbar}(t-\tau)\sqrt{1+q^{2}}}\mbox{sgn}(q)dq\nonumber\\
&=
\begin{pmatrix}
\Theta(\lambda) \\
\Theta(-\lambda) \\
\end{pmatrix}
\frac{im_{0}c}{\pi\hbar}\sqrt{\frac{c}{2}} \int_{0}^{\infty}\frac{\sqrt{q}}{\sqrt[4]{1+q^2}}\sin\left(\frac{m_{0}c}{\hbar}xq\right)\mbox{e}^{-\lambda i\frac{m_{0}c^{2}}{\hbar}(t-\tau)\sqrt{1+q^{2}}}dq\nonumber\\
&=
\begin{pmatrix}
\Theta(\lambda) \\
\Theta(-\lambda) \\
\end{pmatrix}
\frac{im_{0}c}{\pi\hbar}\sqrt{\frac{c}{2}} \int_{0}^{\infty}\frac{\sqrt{q}}{\sqrt[4]{1+q^2}}\sin\left(\frac{m_{0}c}{\hbar}xq\right)\cos\left( \frac{m_{0}c^{2}}{\hbar}(t-\tau)\sqrt{1+q^{2}}\right)dq\nonumber\\
& -i\lambda
\begin{pmatrix}
\Theta(\lambda) \\
\Theta(-\lambda) \\
\end{pmatrix}
\frac{im_{0}c}{\pi\hbar}\sqrt{\frac{c}{2}} \int_{0}^{\infty}\frac{\sqrt{q}}{\sqrt[4]{1+q^2}}\sin\left(\frac{m_{0}c}{\hbar}xq\right)\sin\left( \frac{m_{0}c^{2}}{\hbar}(t-\tau)\sqrt{1+q^{2}}\right)dq\nonumber\\
&\equiv
\begin{pmatrix}
\Theta(\lambda) \\
\Theta(-\lambda) \\
\end{pmatrix}
\frac{im_{0}c}{\pi\hbar}\sqrt{\frac{c}{2}} (f_{1,\tau}^{(-)}(x,t)-i\lambda f_{2,\tau}^{(-)}(x,t))\nonumber\\
\end{align}
where the $f$'s are just the integrals and note that they are independent of the charge signs $\lambda$. We can then calculate the charge density in configuration space by using the $\Phi - x$ version of Eq. (\ref{densphi})
\begin{align}
\rho &= e\Phi_{\lambda,\tau}^{(n)\dagger}(x,t)\sigma_{3}\Phi_{\lambda,\tau}^{(n)}(x,t) \nonumber\\
& = e\frac{m_{0}^{2}c^{3}}{2\pi^{2}\hbar^{2}}
\begin{pmatrix}
\Theta(\lambda) & \Theta(-\lambda) \\
\end{pmatrix}
\begin{pmatrix}
\Theta(\lambda) \\
-\Theta(-\lambda) \\
\end{pmatrix}
\left(|f_{1,\tau}^{(n)}(x,t)|^{2} + |\lambda f_{2,\tau}^{(n)}(x,t)|^{2}\right) \nonumber\\
& = \lambda e \frac{m_{0}^{2}c^{3}}{2\pi^{2}\hbar^{2}} \left(|f_{1,\tau}^{(n)}(x,t)|^{2} + |f_{2,\tau}^{(n)}(x,t)|^{2}\right)
\end{align}
which is clearly positive or negative definite. We can then interpret 
\begin{equation}
P_{\tau}^{(n)}(x,t) = \frac{m_{0}^{2}c^{3}}{2\pi^{2}\hbar^{2}} \left(|f_{1,\tau}^{(n)}(x,t)|^{2} + |f_{2,\tau}^{(n)}(x,t)|^{2}\right)
\end{equation}
as the time evolution of a probability density (in configuration space). Note that $P_{\tau}^{(n)}(x,t)$ is the same for $\lambda = \pm 1$. We now have the means to see how the probability density of a positively or negatively charged particle described by the state $\Phi_{\lambda,\tau}^{(n)}(x,t)$ evolves through time. However, the integrals $f$ strictly diverge so that in order to calculate $P_{\tau}^{(n)}(x,t)$, we need to insert a converging factor $\exp(-\epsilon q)$ and let $\epsilon\rightarrow 0$. In the contour plots, we set $\hbar=c=m_{0}=1$ and choose $\epsilon=0.3$ to be sufficiently small.

\begin{figure}[!htb]
\centering
\includegraphics[width=1\textwidth, height=0.415\textwidth]{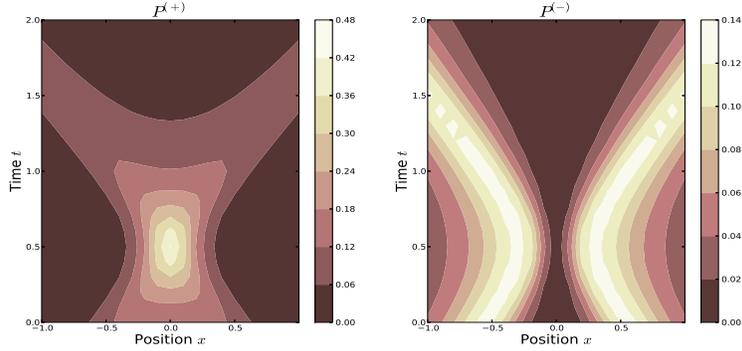}
\caption{Time Evolution of the Probability Density with $\tau = 0.5$}
\label{pdense05}
\end{figure}

\begin{figure}[!htb]
\centering
\includegraphics[width=1\textwidth, height=0.415\textwidth]{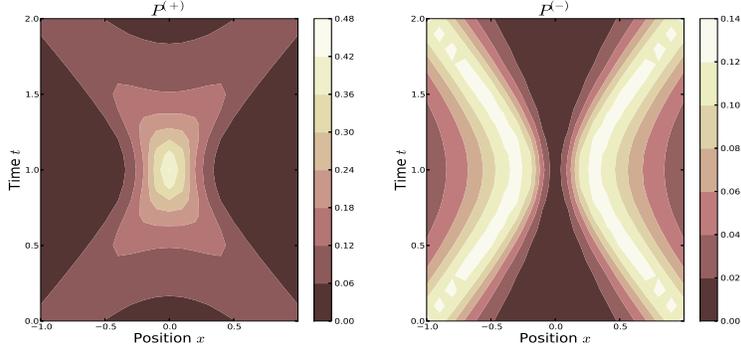}
\caption{Time Evolution of the Probability Density with $\tau = 1.0$}
\label{pdense10}
\end{figure}


As time approaches the eigenvalue $\tau$, it can be observed in the $P^{(+)}$ contour plots that the probability density peaks about the origin, while in the $P^{(-)}$ contour plots, the two probability peaks approach $x=0$, albeit the probability always vanishing at exactly $x=0$. Because of these behaviours, we call $\Phi_{\lambda,\tau}^{(+)}(x,t)$ the non-nodal eigenfunctions and $\Phi_{\lambda,\tau}^{(-)}(x,t)$ the nodal eigenfunctions, in accordance to \cite{gal5,gal6}. The localization observed near $t=\tau$ and $x=0$ can be interpreted as the particle being very much likely to be found at the origin at the time of the eigenvalue. Thus, if we are to accept that the measurement or appearance of a particle at a certain location be characterized by the localization of its wavefunction at that location, then the plots suggest that the particle being described by $\Phi_{\lambda,\tau}^{(n)}(x,t)$ unitarily arrives at the position of the origin at the time of the eigenvalue much like in the works of Galapon et al \cite{gal3}. That is, the particle is in a state of definite arrival time.

\section{The TOA probability distribution $\Pi_{\bar{\phi}_{\lambda}}(\tau)$}
\label{pdist}
We now calculate the time of arrival probability distribution for a particle described by the state $\bar{\Phi}_{\lambda} = (\Theta(\lambda)\;\Theta(-\lambda))^{T} \bar{\phi}_{\lambda}$. That is, we wish to find the probability (density) that the state of the particle will 'collapse' into a state of definite arrival time $\Phi_{\lambda,\tau}^{(n)} = (\Theta(\lambda)\;\Theta(-\lambda))^{T} \phi_{\lambda,\tau}^{(n)}$. Explicitly in momentum representation, $\Pi_{\bar{\phi}_{\lambda}}(\tau)$ would be given by
\begin{align}\label{probdist}
\Pi_{\bar{\phi}_{\lambda}}(\tau) &= \sum_{n} \Pi_{\bar{\phi}_{\lambda}}^{(n)}(\tau) \nonumber\\
&= \sum_{n} \left| \int_{-\infty}^{\infty} \bar{\phi}_{\lambda}^{*}(p) \phi_{\lambda,\tau}^{(n)}(p) dp\right|^{2} \nonumber\\
&= \left| \int_{-\infty}^{\infty} \bar{\phi}_{\lambda}^{*}(p) \sqrt{\frac{c}{4\pi\hbar}} \sqrt{\frac{|p|c}{E_{p}}} \mbox{e}^{\lambda i \tau E_{p}/\hbar} dp\right|^{2}  \nonumber\\
& \;\;\;\;\;\;\;\;\;\; + \left| \int_{-\infty}^{\infty} \bar{\phi}_{\lambda}^{*}(p) \sqrt{\frac{c}{4\pi\hbar}} \sqrt{\frac{|p|c}{E_{p}}} \mbox{e}^{\lambda i \tau E_{p}/\hbar} \mbox{sgn}(p)dp\right|^{2}
\end{align}
We then essentially need to evaluate the overlap integral between the initial state $\bar{\phi}_{\lambda}$ and a TOA eigenfunction $\phi_{\lambda,\tau}^{(n)}$. Note that from section \ref{nonorthsec}, states with $\lambda \neq \lambda'$ have vanishing overlaps. This means that a positively (negatively) charged particle state has a vanishing probability to collapse into a negatively (positively) charged particle state.

As an example, we consider a particle in an initial state given by $\bar{\Phi} = (1 \; 0)^{T} \bar{\phi}$ where
\begin{displaymath}
\bar{\phi}(p) = \frac{1}{\sqrt{m_{0}c\sqrt{\pi/2}}}\exp\left( -\frac{(p-p_{0})^{2}}{m_{0}^{2}c^{2}} -i\frac{x_{0}p}{\hbar}\right)
\end{displaymath}
which can be shown has expectation values of momentum and position $<p> = \left<\bar{\Phi}\left|\hat{p}\bar{\Phi}\right.\right>_{\Phi} = \int_{-\infty}^{\infty}\bar{\Phi}^{\dagger}\sigma_{3}\sigma_{0}p\bar{\Phi}dp = p_{0}$ and $<x> = \left<\bar{\Phi}\left|\hat{q}\bar{\Phi}\right.\right>_{\Phi} = \int_{-\infty}^{\infty}\bar{\Phi}^{\dagger}\sigma_{3}i\hbar\sigma_{0}\frac{\partial}{\partial p}\bar{\Phi}dp = x_{0}$ and spreads in momentum and position $\Delta_{p} = \sqrt{<p^{2}>-<p>^{2}} = \frac{m_{0}c}{2}$ and $\Delta_{x} = \sqrt{<x^{2}>-<x>^{2}} = \frac{\hbar}{m_{0}c}$, respectively. The probability that the particle described by $\bar{\Phi}$ will arrive at the origin during the times $\tau$ and $\tau+d\tau$ is then given by $\Pi_{\bar{\phi}}(\tau)d\tau$ where
\begin{align}
\Pi_{\bar{\phi}}(\tau) &= \left| \int_{-\infty}^{\infty} \frac{1}{\sqrt{m_{0}c\sqrt{\pi/2}}}\mbox{e}^{ -\frac{(p-p_{0})^{2}}{m_{0}^{2}c^{2}} + i\frac{x_{0}p}{\hbar}}\sqrt{\frac{c}{4\pi\hbar}} \sqrt{\frac{|p|c}{E_{p}}} \mbox{e}^{ i \tau E_{p}/\hbar} dp\right|^{2}  \nonumber\\
&\;\;\; + \left| \int_{-\infty}^{\infty} \frac{1}{\sqrt{m_{0}c\sqrt{\pi/2}}}\mbox{e}^{ -\frac{(p-p_{0})^{2}}{m_{0}^{2}c^{2}} + i\frac{x_{0}p}{\hbar}}\sqrt{\frac{c}{4\pi\hbar}} \sqrt{\frac{|p|c}{E_{p}}} \mbox{e}^{ i \tau E_{p}/\hbar} \mbox{sgn}(p)dp\right|^{2}  \nonumber\\
\end{align}
We set $\hbar=c=m_{0}=1$ to illustrate the TOA at the origin probability distribution for a particle with different (average) initial momenta and positions $p_{0}$'s and $x_{0}$'s, respectively. Specifically, we start with a particle moving to the right from the left side of the origin.

\begin{figure}[!htb]
\centering
\includegraphics[width=1\textwidth, height=0.415\textwidth]{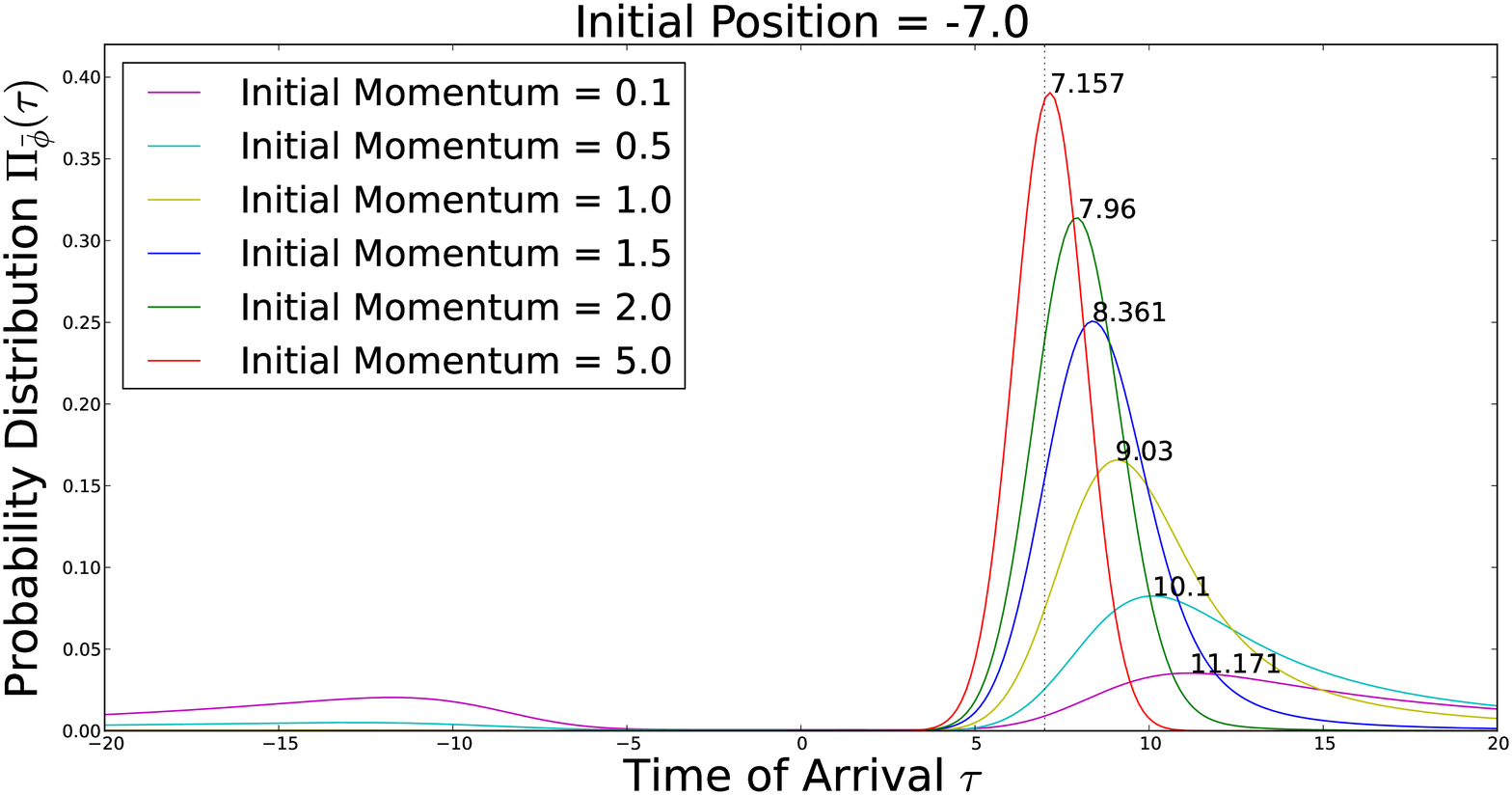}
\caption{TOA Probability Distribution for $x_{0}=-7.0$ and different $p_{0}$'s}
\label{pos2}
\end{figure}

%

\begin{figure}[!htb]
\centering
\includegraphics[width=1\textwidth, height=0.415\textwidth]{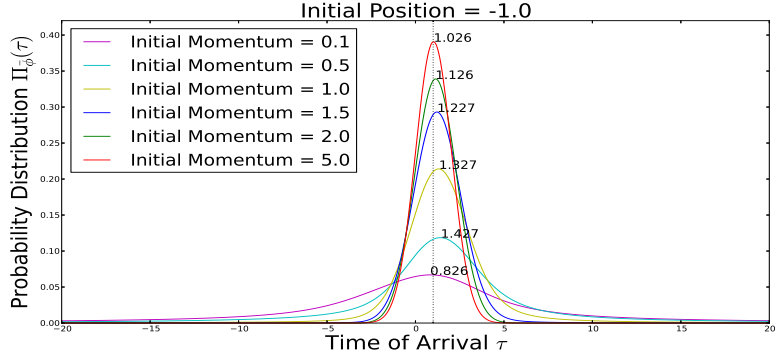}
\caption{TOA Probability Distribution for $x_{0}=-1.0$ and different $p_{0}$'s}
\label{pos5}
\end{figure}


Consider first the cases where the initial average position of the particle $x_{0}$ is sufficiently far from $\Delta_{x}=\frac{\hbar}{m_{0}c}$. One can show that for initial average momenta $p_{0}$ sufficiently far from $\Delta_{p}=\frac{m_{0}c}{2}$, the peaks of $\Pi_{\bar{\phi}}(\tau)$ correspond to $\tau$'s which resemble the free classical relativistic TOA at the origin $t_{class} = -x_{0}\sqrt{p_{0}^{2}+m_{0}^{2}c^{2}}/(p_{0}c)$. That is, the most probable times of arrival at the origin $\tau_{mp}$'s reduce to the $t_{class}$'s. Note also that for these cases, the $\tau_{mp}$'s approach (for increasing $p_{0}$) but are always greater than the TOA at the origin for a photon $t_{ph} = -x_{0}/c$ denoted by the vertical dotted lines in the figures. The time $t_{ph}$ then serves as a limit as $p_{0}$ becomes very large, as expected for a relativistic particle. For being a quantum particle as well, the TOA distribution also have spreads from their respective maxima which may pass through the vertical dotted lines and drop very close to zero as $|\tau|$ becomes very much less than $t_{ph}$. However, this suggests that there is still a non-zero probability for the particle to be superluminal. It can also be seen that the peaks become higher and sharper as $p_{0}$ becomes farther from $\Delta_{p}$ suggesting that the particle becomes more 'classical'. Now consider the cases where $p_{0}$ is sufficiently close $\Delta_{p}$ but with $x_{0}$ still far from $\Delta_{x}$. One can observe in these cases that the peaks have become lower, broader, more flattened, and the $\tau_{mp}$'s become farther from $t_{class}$ suggesting that the quantum behaviour of the particle has become more apparent. Specifically, the probabilities become more evenly distributed, slowly tapering off to larger $\tau$'s, so that the probabilities of the larger times of arrival would become just as probable as $\tau_{mp}$ for $p_{0}\sim 0$, still, in a way, supporting the classical notion that a slow moving particle would take a longer time to reach its destination. In contrast, the probabilities more sharply approach zero for $|\tau|$'s less than $t_{ph}$, suggesting that superluminality of a particle is still less likely. Also, since the momentum probability distribution of the particle can now have a significant spread into momenta with directions opposite to $p_{0}$, the state of the particle can be interpreted as having a rather significant non-zero probability of moving in the opposite direction, reflected in the small probabilities for negative $\tau$'s. That is, there is a non-zero probability that the particle has already arrived at the origin. For the case where $x_{0}$ is sufficiently close to $\Delta_{x}$, the probabilities spread through $t_{ph}$ and into the negative $\tau$'s even for $p_{0}$ far from $\Delta_{p}$. This does not necessarily suggest superluminality, however. This can be taken into account by the position probability distribution having a significant spread at and through the arrival point (the origin), suggesting that the particle has significant probabilities to be initially located to the left, right and even at the origin. Also, when $p_{0}\sim 0$ and $x_{0}\sim 0$, the probabilities become more spread out and $|\tau_{mp}|$ become less than $t_{ph}$. The quantum behaviour is most apparent because of the significant probabilities of the particle moving to the left or right and the particle initially located at the left or right of the origin.

\section{Conclusion}
\label{conc}

In this study, we have derived a maximally symmetric, one-particle TOA operator $\mathcal{\hat{T}}$ canonically conjugate with the Hamiltonian of a free spin-$0$ relativistic charged particle in one spatial dimension. In the non-relativistic limit, we see that (for positive states of the KGE) $\mathcal{\hat{T}}$ reduces to $-m_{0}\hat{T}_{-1,1}$, the TOA operator for a non-relativistic particle with mass $m_{0}$. Being a one-particle operator, it does not mix the positive and negative states of the KGE. That is, if we start with a particle, we would still end up with the same one. It then makes sense to consider systems starting with a relativistic particle in some initial state $\Phi_{0}$, and asking for, say, its average TOA at the the origin $\left<\Phi_{0}\left|\mathcal{\hat{T}}\Phi_{0}\right.\right>_{\Phi}$ which, as we have seen, is a real number (as desired for expectation values). We also solved for the eigenfunctions of $\mathcal{\hat{T}}$ (which form a complete and non-orhtogonal set and reduces to the appropriate non-relativistic limit apart from an irrelevant phase factor), and investigated their dynamical behaviour by considering the time evolution of their associated probability densities $P_{\tau}^{(n)}(x,t)$ (which is non-negative as expected). Note here that $P_{\tau}^{(n)}(x,t)$ is the same whether we are considering a positively charged or negatively charged particle. We also saw from the plots of $P_{\tau}^{(n)}(x,t)$ that the eigenfunctions become more localized at the origin at their corresponding eigenvalues. This suggests that a particle in a state $\Phi_{\lambda,\tau}^{(n)}$ is in a state of definite arrival time $\tau$ at the origin as what we should expect for the eigenfunctions of a TOA operator. We also calculated for the probability distribution $\Pi_{\bar{\phi}_{\lambda}}(\tau)$. That is, we calculated the probability (density) that a particle in some initial state would arrive at the origin at time $\tau$. As an example, we let the initial state of a particle be normally distributed about some various average initial positions and momenta, $x_{0}$'s and $p_{0}$'s, respectively, and plot their corresponding probability densities. As expected, the plots revealed relativistic classical behaviour when $x_{0}$ and $p_{0}$ are sufficiently far from their respective spreads, while quantum wave-like behaviour became apparent when either $x_{0}$ or $p_{0}$ are close to their respective spreads. 

These results give us confidence that we are on the correct path in promoting time to an observable in relativistic quantum mechanics. This may be in a step forward towards having a formalism where physical observables are on equal footing, perhaps paving the way in reconciling the inconsistent notions of time in quantum mechanics and general relativity. As for some speculative examples, upon decomposing or foliating spacetime into spacelike hypersurfaces and letting them evolve through a certain time variable which leads to the Wheeler-DeWitt equations \cite{qg1}, perhaps one can define appropriate 'momentum and energy' operators for the gravitational field and impose that they should be canonically conjugate to some 'space and time' operators whose eigenvalues may be the spacetime points. Although interpretations of such operators should be studied further. Elsewhere, in Loop Quantum Gravity, the discreteness of space was implied through the volume operator \cite{qg2}. Similarly, perhaps a time operator, conjugate to a certain Hamiltonian, can also be constructed which may imply the discreteness of time. Or perhaps there is a spacetime volume operator so that there may be a smallest chunk of spacetime. Also, in the continuum 1+1 dimensional model of Causal Dynamical Triangulations, one can calculate a Hamiltonian governing the (proper) time evolution of a universe of some initial size or length \cite{qg3, qg4, qg5}. As the spatial length of the universe is considered as an observable in the formalism, perhaps the temporal length of the universe may also be considered as an observable which may be represented by an operator conjugate to the said Hamiltonian. 

Future studies may include the interacting case. Constructing one-particle operators may be difficult, however. This is because separating the equations of motion for positive and negative states may not be possible. In the picture provided by QFT, the interacting fields suggest that the particles associated with these fields collide and interact - creating and annihilating particles and anti-particles. That is, the non-conservation of particle number may be necessary. Thus, restricting our picture to one particle would be more difficult. Along the lines of the interacting case, a similar system may be formulated in a curved spacetime. The formulation of the equations of motion (the KGE in this case) should then be able to accomodate this nontrivial curvature. Another area would be to relate possible results here to QFT since they should be able to describe the same systems. As an example, both approaches should be able to provide a probability amplitude for a particle to start from an initial spacetime point and arrive to a final spacetime point. Also, since the derivations in this study is formal, further studies on the domain of $\mathcal{\hat{T}}$ is needed. We should be able to define a dense domain on which $\mathcal{\hat{T}}$ can act upon for it to be a meaningful quantum operator. Lastly, this study focuses on the spin-$0$ particle. We should also be able to construct a one-particle TOA operator for particles with other spins (i.e. spin-$1/2$) canonically conjugate with its Hamiltonian.

\appendix
\section{Some useful relations}

Consider the commutator between the momentum and position operators $\hat{p}$ and $\hat{q}$, respectively, $\left[\hat{p},\hat{q}\right] = -i\hbar$. We calculate the following:

\begin{align}
\left[\hat{p},\hat{q}^{k}\right] &=\left[\hat{p},\hat{q}^{k-1}\hat{q}\right] \nonumber\\
&=\hat{q}^{k-1}\left[\hat{p},\hat{q}\right]+\left[\hat{p},\hat{q}^{k-1}\right]\hat{q} \nonumber\\
&=-i\hbar\hat{q}^{k-1}+\left[\hat{p},\hat{q}^{k-2}\hat{q}\right]\hat{q} \nonumber\\
&=-i\hbar\hat{q}^{k-1}+(\hat{q}^{k-2}\left[\hat{p},\hat{q}\right]+\left[\hat{p},\hat{q}^{k-2}\right]\hat{q})\hat{q} \nonumber\\
&=-2i\hbar\hat{q}^{k-1}+\left[\hat{p},\hat{q}^{k-2}\right]\hat{q}^{2} \nonumber\\
& \;\;\vdots \nonumber\\
&=-i\hbar k\hat{q}^{k-1} \nonumber\\
\left[\hat{p}^{2},\hat{q}^{k}\right] &= \hat{p}\left[\hat{p},\hat{q}^{k}\right] + \left[\hat{p},\hat{q}^{k}\right]\hat{p}\nonumber\\
&= -i\hbar k\hat{p}\hat{q}^{k-1}+ \left[\hat{p},\hat{q}^{k}\right]\hat{p}\nonumber\\
&= -i\hbar k(\hat{q}^{k-1}\hat{p} - i\hbar (k-1)\hat{q}^{k-2}) + \left[\hat{p},\hat{q}^{k}\right]\hat{p}\nonumber\\
&= -2i\hbar k\hat{q}^{k-1}\hat{p} -\hbar^{2}k(k-1)\hat{q}^{k-2}\nonumber\\
&= \hat{p}\left[\hat{p},\hat{q}^{k}\right] -i\hbar k\hat{q}^{k-1}\hat{p}\nonumber\\
&= \hat{p}\left[\hat{p},\hat{q}^{k}\right] -i\hbar k(\hat{p}\hat{q}^{k-1} + i\hbar (k-1)\hat{q}^{k-2}) \nonumber\\
&= -2i\hbar k\hat{p}\hat{q}^{k-1} + \hbar^{2}k(k-1)\hat{q}^{k-2}\nonumber\\
\end{align}
Now, consider the complete and linearly independent set of Bender-Dunne operators denoted by $\hat{T}_{m,n}$. These are the Weyl ordered quantization of the classical monomial $p^{m}q^{n}$
\begin{align}
\hat{T}_{m,n} &=\frac{1}{2^n}\sum_{k=0}^{\infty} \frac{n!}{k!(n-k)!} \hat{q}^{k} \hat{p}^{m} \hat{q}^{n-k} \nonumber\\
&=\frac{1}{2^n}\sum_{k=0}^{\infty} \frac{n!}{k!(n-k)!} \hat{q}^{n-k} \hat{p}^{m} \hat{q}^{k} \nonumber\\
\end{align}
and calculate the following:
\begin{align}
\hat{p}^{2}\hat{T}_{m,n} &= \frac{1}{2^n}\sum_{k=0}^{\infty} \frac{n!}{k!(n-k)!} \hat{p}^{2}\hat{q}^{k} \hat{p}^{m} \hat{q}^{n-k} \nonumber\\
&= \frac{1}{2^n}\sum_{k=0}^{\infty} \frac{n!}{k!(n-k)!} (\hat{q}^{k}\hat{p}^{2} -2i\hbar k\hat{q}^{k-1}\hat{p} -\hbar^{2}k(k-1)\hat{q}^{k-2}) \hat{p}^{m} \hat{q}^{n-k} \nonumber\\
&= \hat{T}_{m+2,n} - i\hbar n\hat{T}_{m+1,n-1} -\frac{\hbar^{2}n(n-1)}{4}\hat{T}_{m,n-2}\nonumber\\
\hat{T}_{m,n}\hat{p}^{2} &= \frac{1}{2^n}\sum_{k=0}^{\infty} \frac{n!}{k!(n-k)!} \hat{q}^{n-k} \hat{p}^{m} \hat{q}^{k}\hat{p}^{2} \nonumber\\
&= \frac{1}{2^n}\sum_{k=0}^{\infty} \frac{n!}{k!(n-k)!}\hat{q}^{n-k}\hat{p}^{m} (\hat{p}^{2}\hat{q}^{k} +2i\hbar k\hat{p}\hat{q}^{k-1} - \hbar^{2}k(k-1)\hat{q}^{k-2}) \nonumber\\
&= \hat{T}_{m+2,n} + i\hbar n\hat{T}_{m+1,n-1} -\frac{\hbar^{2}n(n-1)}{4}\hat{T}_{m,n-2}\nonumber\\
\end{align}
Consider also the linearly independent, complete set of $2$x$2$ matrices, denoted by
\begin{displaymath}
\sigma_{0} = 
\begin{pmatrix}
 1 & 0\\
 0 & 1 \\
\end{pmatrix},
\sigma_{1} = 
\begin{pmatrix}
 0 & 1\\
 1 & 0 \\
\end{pmatrix},
\sigma_{2} = 
\begin{pmatrix}
 0 & -i\\
 i & 0 \\
\end{pmatrix},
\sigma_{3} = 
\begin{pmatrix}
 1 & 0\\
 0 & -1 \\
\end{pmatrix}
\end{displaymath}
Note that $\sigma_{0}$ is just the identity matrix. Using the property of the last three matrices $\sigma_{x}\sigma_{y}=i\sigma_{z}$ where, $(x,y,z)$ are just cyclic permutations of $(1,2,3)$, and $\sigma_{j}\sigma_{j} = \sigma_{0}$ for $j=0,1,2,3$, we arrive at some relevant commutators and anti-commutators of the matrices $\sigma_{j}$. Namely,
\begin{align}
[\sigma_{3},\sigma_{j}] &= 2i\delta_{j,1}\sigma_{2} - 2i\delta_{j,2}\sigma_{1} \nonumber\\
[\sigma_{3}+i\sigma_{2},\sigma_{j}] &= 2\delta_{j,1}(\sigma_{3}+i\sigma_{2}) -2i\delta_{j,2}\sigma_{1} - 2\delta_{j,3}\sigma_{1}\nonumber\\
\{\sigma_{3}+i\sigma_{2},\sigma_{j}\} &= 2\delta_{j,0}(\sigma_{3}+i\sigma_{2}) + 2i\delta_{j,2}\sigma_{0}+ 2\delta_{j,3}\sigma_{0}\nonumber\\
\end{align}
Lastly, consider the transform $U$ and its inverse $U^{-1}$ in momentum representation.
\begin{equation}
U^{\pm 1} = \frac{(m_{0}c^{2}+E_{p})\sigma_{0} \mp (m_{0}c^{2}-E_{p})\sigma_{1}}{\sqrt{4m_{0}c^{2}E_{p}}}
\end{equation}
where, $E_{p} = \sqrt{p^{2}c^{2}+m_{0}^{2}c^{4}}$ and calculate
\begin{align}
\frac{\partial}{\partial p} E_{p} &= \frac{pc^{2}}{E_{p}}\nonumber\\
\frac{\partial}{\partial p} U^{-1} &= \frac{m_{0}c^{2}(\sigma_{0}+\sigma_{1})}{\sqrt{4m_{0}c^{2}}}\frac{\partial}{\partial p} \left(E_{p}^{-1/2}\right) + \frac{(\sigma_{0}-\sigma_{1})}{\sqrt{4m_{0}c^{2}}}\frac{\partial}{\partial p} \left(E_{p}^{1/2}\right)\nonumber\\
&= \frac{m_{0}c^{2}(\sigma_{1}^{2}+\sigma_{1})}{\sqrt{4m_{0}c^{2}}} \left(-\frac{1}{2}E_{p}^{-3/2}\frac{pc^{2}}{E_{p}}\right) + \frac{(\sigma_{1}^{2}-\sigma_{1})}{\sqrt{4m_{0}c^{2}}} \left(\frac{1}{2}E_{p}^{-1/2}\frac{pc^{2}}{E_{p}}\right)\nonumber\\
&= \left(\frac{m_{0}c^{2}(\sigma_{1}+\sigma_{0})}{\sqrt{4m_{0}c^{2}E_{p}}}-\frac{(\sigma_{1}-\sigma_{0})E_{p}}{\sqrt{4m_{0}c^{2}E_{p}}} \right)\left(-\frac{pc^{2}}{2E_{p}^{2}}\right)\sigma_{1}\nonumber\\
&= U^{-1}\left(-\frac{pc^{2}}{2E_{p}^{2}}\right)\sigma_{1}\nonumber\\
U\sigma_{3}U^{-1} &= \frac{\sigma_{3}(m_{0}c^{2}+E_{p})^{2}+2i\sigma_{2}(m_{0}^{2}c^{4}-E_{p}^{2})+\sigma_{3}(m_{0}c^{2}-E_{p})^{2}}{4m_{0}c^{2}E_{p}}\nonumber\\
&= \sigma_{3}\frac{2m_{0}^{2}c^{4}+p^{2}c^{2}}{2m_{0}c^{2}E_{p}} - i\sigma_{2}\frac{p^{2}c^{2}}{2m_{0}c^{2}E_{p}}\nonumber\\
&= \sigma_{3}\frac{m_{0}c^{2}}{E_{p}} + (\sigma_{3}-i\sigma_{2})\frac{p^{2}}{2m_{0}E_{p}}\nonumber\\
U(\sigma_{3}+i\sigma_{2})U^{-1} &=U\sigma_{3}U^{-1}+ \frac{i\sigma_{2}(m_{0}c^{2}+E_{p})^{2}+2\sigma_{3}(-p^{2}c^{2})+i\sigma_{2}(m_{0}c^{2}-E_{p})^{2}}{4m_{0}c^{2}E_{p}}\nonumber\\
&=U\sigma_{3}U^{-1}-\sigma_{3}\frac{p^{2}}{2m_{0}E_{p}}+i\sigma_{2}\frac{2m_{0}^{2}c^{4}+p^{2}c^{2}}{2m_{0}c^{2}E_{p}}\nonumber\\
&= (\sigma_{3}+i\sigma_{2})\frac{m_{0}c^{2}}{E_{p}}\nonumber\\
\end{align}
We can then calculate the Klein-Gordon Hamiltonian in the $\Phi - p$ representation $\mathcal{\hat{H}}_{\Phi}\Phi = U\mathcal{\hat{H}}_{\Psi}U^{-1}\Phi$.
\begin{align}
\mathcal{\hat{H}}_{\Phi}\Phi &= U\left((\sigma_{3}+i\sigma_{2})\frac{p^{2}}{2m_{0}} + \sigma_{3}m_{0}c^{2}\right)U^{-1}\Phi\nonumber\\
&= U(\sigma_{3}+i\sigma_{2})U^{-1}\frac{p^{2}}{2m_{0}}\Phi + U\sigma_{3}U^{-1} m_{0}c^{2}\Phi\nonumber\\
&=(\sigma_{3}+i\sigma_{2})\frac{m_{0}c^{2}}{E_{p}}\frac{p^{2}}{2m_{0}}\Phi + \left(\sigma_{3}\frac{m_{0}c^{2}}{E_{p}} + (\sigma_{3}-i\sigma_{2})\frac{p^{2}}{2m_{0}E_{p}}\right)m_{0}c^{2}\Phi\nonumber\\
&=\sigma_{3}\frac{p^{2}c^{2}+m_{0}^{2}c^{4}}{E_{p}}\Phi\nonumber\\
&= \sigma_{3}E_{p}\Phi
\end{align}
Note that the Hamiltonian is even (i.e. it does not mix positive and negative states)

\section{Eigenfunctions in the $\Psi - x$ representation}
Using Eq. (\ref{eigftq}), we transform it back to the Schrodinger ($\Psi$) representation by $\Psi_{\lambda, \tau}^{(n)}(q,t) = U^{-1}\Phi_{\lambda, \tau}^{(n)}(q,t)$ where,
\begin{align}
U^{- 1} &= \frac{(m_{0}c^{2}+E_{p})\sigma_{0} + (m_{0}c^{2}-E_{p})\sigma_{1}}{\sqrt{4m_{0}c^{2}E_{p}}}\nonumber\\
&= \frac{(1+\sqrt{1+q^2})\sigma_{0} + (1-\sqrt{1+q^2}) \sigma_{1}}{2\sqrt[4]{1+q^2}}\nonumber\\
\end{align}
In the $\Psi - p$ representation, the eigenfunctions then take the form
\begin{align}
\Psi_{\lambda, \tau}^{(+)}(q,t) &= \sqrt{\frac{c}{\pi\hbar}}\frac{\sqrt{|q|}}{4\sqrt{1+q^2}}e^{-\lambda i\frac{m_{0}c^{2}}{\hbar}(t-\tau)\sqrt{1+q^{2}}}
\begin{pmatrix}
1+\lambda\sqrt{1+q^2} \\
1-\lambda\sqrt{1+q^2} \\
\end{pmatrix}
\nonumber\\
&= \sqrt{\frac{c}{\pi\hbar}}\frac{\sqrt{|q|}}{4}e^{-\lambda i\frac{m_{0}c^{2}}{\hbar}(t-\tau)\sqrt{1+q^{2}}}
\begin{pmatrix}
(1+q^2)^{-1/2}+\lambda \\
(1+q^2)^{-1/2}-\lambda \\
\end{pmatrix}
\nonumber\\
\Psi_{\lambda, \tau}^{(-)}(q,t) &= \sqrt{\frac{c}{\pi\hbar}}\frac{\sqrt{|q|}}{4\sqrt{1+q^2}}e^{-\lambda i\frac{m_{0}c^{2}}{\hbar}(t-\tau)\sqrt{1+q^{2}}}
\begin{pmatrix}
1+\lambda\sqrt{1+q^2} \\
1-\lambda\sqrt{1+q^2} \\
\end{pmatrix}
\mbox{sgn}(q)
\nonumber\\
&= \sqrt{\frac{c}{\pi\hbar}}\frac{\sqrt{|q|}}{4}e^{-\lambda i\frac{m_{0}c^{2}}{\hbar}(t-\tau)\sqrt{1+q^{2}}}
\begin{pmatrix}
(1+q^2)^{-1/2}+\lambda \\
(1+q^2)^{-1/2}-\lambda \\
\end{pmatrix}
\mbox{sgn}(q)
\nonumber\\
\end{align}
To obtain the eigenfunctions in position representation, we take its Fourier Transform. Namely,
\begin{align}
\Psi_{\lambda, \tau}^{(n)}(x,t) &= \frac{1}{\sqrt{2\pi\hbar}}\int_{-\infty}^{\infty} e^{ixp/\hbar}\Psi_{\lambda, \tau}^{(n)}(p,t) dp \nonumber\\
&= \frac{m_{0}c}{\sqrt{2\pi\hbar}}\int_{-\infty}^{\infty} e^{im_{0}cxq/\hbar}\Psi_{\lambda, \tau}^{(n)}(q,t) dq \nonumber\\
\end{align}
Explicitly, the eigenfunctions in $\Psi - x$ representation are
\small
\begin{align}
\Psi_{\lambda, \tau}^{(+)}(x,t) &= \frac{m_{0}c^{3/2}}{2^{5/2}\pi\hbar}\int_{-\infty}^{\infty} e^{i\frac{m_{0}c}{\hbar}xq}\sqrt{|q|}e^{-\lambda i\frac{m_{0}c^{2}}{\hbar}(t-\tau)\sqrt{1+q^{2}}}
\begin{pmatrix}
(1+q^2)^{-1/2}+\lambda \\
(1+q^2)^{-1/2}-\lambda \\
\end{pmatrix}
dq
\nonumber\\
&= \frac{m_{0}c^{3/2}}{2^{3/2}\pi\hbar}\int_{0}^{\infty} \cos\left(\frac{m_{0}c}{\hbar}xq\right)\sqrt{q}e^{-\lambda i\frac{m_{0}c^{2}}{\hbar}(t-\tau)\sqrt{1+q^{2}}}
\begin{pmatrix}
(1+q^2)^{-\frac{1}{2}}+\lambda \\
(1+q^2)^{-\frac{1}{2}}-\lambda \\
\end{pmatrix}
dq
\nonumber\\
\Psi_{\lambda, \tau}^{(-)}(x,t) &= \frac{m_{0}c^{3/2}}{2^{5/2}\pi\hbar}\int_{-\infty}^{\infty} e^{i\frac{m_{0}c}{\hbar}xq}\sqrt{|q|}e^{-\lambda i\frac{m_{0}c^{2}}{\hbar}(t-\tau)\sqrt{1+q^{2}}}
\begin{pmatrix}
(1+q^2)^{-1/2}+\lambda \\
(1+q^2)^{-1/2}-\lambda \\
\end{pmatrix}
\mbox{sgn}(q)dq
\nonumber\\
&= \frac{m_{0}c^{3/2}i}{2^{3/2}\pi\hbar}\int_{0}^{\infty} \sin\left(\frac{m_{0}c}{\hbar}xq\right)\sqrt{q}e^{-\lambda i\frac{m_{0}c^{2}}{\hbar}(t-\tau)\sqrt{1+q^{2}}}
\begin{pmatrix}
(1+q^2)^{-\frac{1}{2}}+\lambda \\
(1+q^2)^{-\frac{1}{2}}-\lambda \\
\end{pmatrix}
dq
\nonumber\\
\end{align}

%

\end{document}